\documentclass[preprint,3p,onecolumn]{elsarticle}
\usepackage{graphicx}
\usepackage{amsmath}
\usepackage{amssymb}
\usepackage{bbm}
\usepackage{xspace}
\usepackage{color}
\usepackage{footnote}
\usepackage{comment}

\usepackage{epstopdf}

\usepackage[caption=false]{subfig}
\usepackage{hyperref}

\usepackage{pseudocode}

\definecolor{gray}{RGB}{150,150,150}
\definecolor{green3}{RGB}{00,200,00}

\newcommand{\appropto}{\mathrel{\vcenter{
  \offinterlineskip\halign{\hfil$##$\cr
    \propto\cr\noalign{\kern2pt}\sim\cr\noalign{\kern-2pt}}}}}

\DeclareMathOperator*{\SumIntx}{%
\mathchoice%
  {\ooalign{$\displaystyle\sum$\cr\hidewidth$\displaystyle\int$\hidewidth\cr}}
  {\ooalign{\raisebox{.14\height}{\scalebox{.7}{$\textstyle\sum$}}\cr\hidewidth$\textstyle\int$\hidewidth\cr}}
  {\ooalign{\raisebox{.2\height}{\scalebox{.6}{$\scriptstyle\sum$}}\cr$\scriptstyle\int$\cr}}
  {\ooalign{\raisebox{.2\height}{\scalebox{.6}{$\scriptstyle\sum$}}\cr$\scriptstyle\int$\cr}}
}
\newcommand{\SumInt}[1]{\SumIntx\limits_{#1}}

\newcommand{\tbd}[1]{}
\newcommand{\betaJ}{\mathbbm{b}}
\newcommand{\BK}{{\cal I}}
\newcommand{\OLD}[1]{}
\newcommand{\NEW}[1]{}
\newcommand{\mn}[1]{}
\newcommand{\mncomm}[1]{}
\newcommand{\ad}[1]{}
\newcommand{\adcomm}[1]{}
\newcommand{\ea}[1]{}
\newcommand{\eacomm}[1]{}
\newcommand{\wvlcomm}[1]{}

\newcommand{\fig}[1]{Fig.\thinspace{}\ref{#1}}

\newcommand{\Fig}[1]{Fig.\thinspace{}\ref{#1}}
\newcommand{\eq}[1]{Eq.\thinspace{}(\ref{#1})}
\newcommand{\Eq}[1]{Eq.\thinspace{}(\ref{#1})}

\newcommand{\Se}{Sec.\@\xspace}

\newcommand{\avg}[1]{\left\langle #1 \right\rangle}

\newcommand{\etal}[0]{\textit{et al.}}

\newcommand{\beq}{\begin{equation}}
\newcommand{\eeq}{\end{equation}}

\newcommand{\myhat}[1]{#1^{*}}
\newcommand{\vvmyhat}[1]{{\vv #1}^{*}}

\newcommand{\vv}[1]{\boldsymbol{#1}}

\newcommand{\Npair}{N_\text{p}}
\newcommand{\CS}{\,\boldsymbol \vert\,}

\biboptions{longnamesfirst,sort&compress}

\begin{document}

\begin{frontmatter}

\title{Nested sampling, statistical physics and the Potts model}

\author[oeaw]{Manuel J. Pfeifenberger\corref{cor1}}
\address[oeaw]{Erich Schmid Institute of Materials Science, Austrian Academy of Sciences, 8700 Leoben, Austria}
\cortext[cor1]{Corresponding author}
\ead{manuel.pfeifenberger@oeaw.ac.at}
\author[itp]{Michael Rumetshofer}
\author[itp]{Wolfgang von der Linden}
\address[itp]{Institute of Theoretical and Computational Physics, Graz University of Technology, 8010 Graz, Austria}

\date{\today}

\begin{abstract}
We present a systematic study of the nested sampling algorithm based on the example of the Potts model. This model, which exhibits a first order phase transition  for $q>4$, exemplifies a generic numerical challenge in statistical physics: The evaluation of the partition function and thermodynamic observables, which involve high dimensional  sums of sharply structured multi-modal density functions. It poses a major  challenge to most standard numerical techniques, such as Markov Chain Monte 
Carlo. 
In this paper we will demonstrate that nested sampling is particularly suited for such problems and
it has a couple of advantages. For calculating the partition function 
of the Potts model with $N$ sites: a) one run stops after $O(N)$
moves, so it takes $O(N^{2})$ operations for the run, b) only a single run is required to compute the partition function along with the assignment of confidence intervals,
c) the confidence intervals of the logarithmic partition function decrease with $1/\sqrt{N}$ and d)
a single run allows to compute quantities for all temperatures while 
the autocorrelation time is very small, irrespective of  temperature.
Thermodynamic expectation values of observables, which are completely determined by the bond configuration in the representation of Fortuin and
Kasteleyn, 
like the Helmholtz free energy, the internal energy as well as the entropy and heat capacity, can be calculated in the same single run needed for the partition function along with their confidence intervals.
In contrast, thermodynamic expectation values of magnetic properties like the magnetization and the magnetic susceptibility require sampling the additional spin degree of freedom.
Results and performance are studied in detail and compared with those obtained with multi-canonical sampling. 
Eventually the implications of the findings on a parallel implementation of nested sampling are outlined.

\end{abstract}



\end{frontmatter}

\section{Introduction}\label{sec:introduction}
Monte Carlo (MC) simulations are the most important instrument for the evaluation of integrals or sums in a high dimensional phase space. When it comes to the computation of 
partition functions there are only very few 
reliable techniques, such as simulated tempering \cite{landau_guide_2009}, 
multi-canonical sampling (MUCA) \cite{berg_introduction_1999,berg_multicanonical_2003}, or
multi-bondic cluster algorithms (MUBO) \cite{janke_multibondic_1995} available that guarantee perfect mixing.
Skilling, however, gave an example, where also MUCA would fail \cite{skilling_nested_2006}.
Basically, these algorithms  are trying to enhance the efficiency of the MC algorithm by
 flattening  the probability  distribution.
A conceptual completely different approach, named \emph{nested sampling} (NESA), has been suggested by Skilling  \cite{skilling_nested_2006}.
It is a promising way for estimating high dimensional,
multi-modal integrals or sums and is based on a mapping to a  Lebesgue integral along with a 
novel sampling technique.
Since its development it has already found its way into various fields of research. Especially  in
 statistics and Bayesian inference various applications already exist 
\cite{aitken_nested_2013,burkoff_exploring_2012,feroz_exploring_2013,mukherjee_nested_2006,qiao_generalized_2014}. 
Furthermore, for atomistic modelling and materials simulations NESA has proven to be an highly efficient algorithm \cite{partay_efficient_2010,partay_nested_2014,baldock_determining_2016}.
A first, though incomplete, application in the field of statistical physics, in particular for the Potts model, has been presented by Murray \etal  \cite{murray_nested_2005}.
The Potts model provides, despite of its simple structure, a wide variety of 
interesting physical  properties and the availability of exact results for  certain 
quantities of the two dimensional model makes it an optimal playground for testing 
new approaches in simulation techniques.
For parameters, where the model exhibits a first order phase transition, the numerical evaluation of the partition function poses a severe
difficulty for standard MC algorithms, because
at first order phase transitions the autocorrelation times can become huge. Moreover, such systems are 
characterized by a double-peak structure in the probability density for the energy with a pronounced minimum in between, which is exponentially suppressed due to the interface tension. This causes severe mixing problems in standard Monte Carlo techniques \cite{Janke_2008}.
In this case two or more well separated phase space regions need to be explored.
Transitions between these regions are possible but
 very improbable and therefore the relative weights of the maxima will not be  sampled correctly. 
 Multi-canonical sampling (MUCA) and multi-bondic sampling (MUBO) 
 are particularly tailored to overcome this barrier \cite{berg_introduction_1999,berg_multicanonical_2003,janke_multibondic_1995}.
 A detailed analysis and comparison is given by Janke \cite{1998PhyA..254..164J}.
 
In the present work the implementation of nested sampling for evaluating the partition function and thermodynamic expectation values of the Potts model and its accuracy are significantly improved as compared to that in \cite{murray_nested_2005} and in particular, no additional thermodynamic integration is required.
 We compute thermodynamic expectation values like the internal energy, the entropy, the magnetization and the magnetic susceptibility along with confidence intervals applying NESA, discuss the results and compare to MUCA, theoretical values and limiting cases.
Moreover, the performance of NESA is thoroughly investigated and compared with the performance of MUCA.
Already in his original paper \cite{skilling_nested_2006} Skilling mentioned the possibility of a faster exploration of the phase space via a parallel implementation of nested sampling.
This idea has recently been picked up in various publications 
(\cite{burkoff_exploring_2012,henderson_parallelized_2014,martiniani_superposition_2014,vanderbauwhede_implementing_2013}).
In this paper we will explore  the capability of parallel nested sampling.

The paper is organized as follows:
In \Se\ref{sec:pottsmodel} the  Potts model is introduced. 
The MC methods employed by us to compute the partition function of the Potts model, namely multi-canonical simulation and  nested sampling are described in \Se\ref{ssec:muca} and \Se\ref{ssec:nesa}, respectively.
Being in the focus of our investigation, NESA is treated in a more elaborate way.
Results for the Potts model are compiled in \Se\ref{ssec:nsbasic}. A performance comparison of NESA and MUCA is shown in \Se\ref{sec:compnsti}. Furthermore, it is presented how thermodynamic variables can be evaluated from a single  nested sampling run. 
Consequences for a parallel implementation are discussed in \Se\ref{ssec:parallel_ns}.
Finally the results are discussed and potential generalizations are outlined.

\section{Potts model}\label{sec:pottsmodel}

One of the most investigated models in statistical physics is the Potts model \cite{wu_potts_1982}.
The corresponding Hamiltonian, without  external fields, reads 
\begin{align}\label{eq:potts_energy}
	H (\mathbf{s}) &= - J\:\sum_{\left\langle i,j \right\rangle} \big(\delta_{s_{i},s_{j}}-1\big)
	= -J \big(N_\text{eq}(\mathbf{s})	-\Npair\big)\;.
\end{align}
The dynamic variables $s_{i}$ of the model, referred to as spin or color, can assume the integer values between 1 and $q$.  
The term $N_\text{eq}(\mathbf{s}) = \sum_{\left\langle i,j \right\rangle} \delta_{s_{i},s_{j}}$
is the number of nearest neighbour pairs with equal spin, and 
$\Npair$ is the total number of nearest neighbour pairs in the lattice under consideration.
Here the exchange coupling $J$ is positive and site independent. The lattice indices are denoted by  $i$ and $j$, ranging from 1 to the number of sites $N$.
 The sum only includes nearest neighbour interactions which is denoted by $\left\langle i,j \right\rangle$. 
The sought-for partition function reads
\begin{align} \label{eq:partition:fct:spin}
Z_{\text{P}}(\betaJ ) &= \sum_{\mathbf{s}} e^{\betaJ\:\sum_{\left\langle i,j \right\rangle} \big(\delta_{s_{i},s_{j}}-1\big)}\;,
\end{align} 
with the abbreviation $\betaJ = \beta J$.
Two limiting cases can easily be determined
\begin{align} \label{eq:limits}
Z(\betaJ)&\underset{\betaJ\to 0}{\longrightarrow} q^N\;,\qquad
Z(\betaJ)\underset{\betaJ\to \infty}{\longrightarrow} q\;,
\end{align} 
which will be of interest later on. 
In the first case (high temperature limit), the Boltzmann factor is one for all spin configurations while in the second (low temperature limit) only those configurations contribute, where all spins have the same value.
Our investigations are restricted to 2d square lattices with periodic boundary conditions. 
The infinite square lattice exhibits for $q\leq4$ ($q > 4$) a second (first) order  phase transition.
The exact critical inverse temperature for the Potts model on a two dimensional infinite square lattice follows from self-duality of the low and high temperature region \cite{baxter_potts_1973}
\begin{align} \label{eq:Tc:potts}
\betaJ_{c} = \ln (1+ \sqrt{q}).
\end{align}

\section{Evaluation of the partition function}\label{sec:evaluation_partition}

Quite generally, the partition function $Z$, associated to the thermodynamic potential (Helmholtz free energy), contains the entire thermodynamic information of a system in the canonical ensemble.
It can be expressed as
\begin{eqnarray}\label{eq:thermoint0}
    Z = \SumInt {\mathbf{x}}\: L(\mathbf{x})\:\pi(\mathbf{x}),
\end{eqnarray} 
where $\mathbf{x}$ describes a point in a multidimensional phase space,
which can either be continuous or discrete.
In the following $L(\mathbf{x})$ is
 denoted as likelihood function and $\pi(\mathbf{x})$ as prior probability.
Assuming the likelihood function $L(\mathbf{x})$ shows a strong variation, then a classical 
Markov Chain MC needs a huge sample size to yield reasonable variances \cite{von_toussaint_bayesian_2011}.
For the evaluation of Z special methods exist. Two of them will be presented in the following subsections.

\subsection{Multi-canonical simulation and the partition function}\label{ssec:muca}

 Multi-canonical simulation (MUCA) is a method introduced by Berg and Neuhaus \cite{Berg_MUCA_1991,Berg_1992} and thoroughly investigated by Janke \cite{Janke_1998,Janke_2008}.
 The underlying idea is briefly as follows.
We start out  from the Boltzmann distribution 
 \begin{align}
  p_{\text{can}}(\mathbf{x}) = \frac{1}{Z} e^{-\betaJ E(\mathbf{x})} \;,
 \end{align}
where  energies are measured in units of $J$.
 For the multi-canonical probability distribution an additional weight function 
 $w(E):=e^{-g(E)}$ is introduced that 
defines a new probability distribution
 \begin{align}
  p_{\text{muca}}(\mathbf{x}) = p_{\text{can}}(\mathbf{x}) w(E(\mathbf{x})),
 \end{align}
which is used in the MC simulation.
 The goal of MUCA is to choose the  additional weights such that the resulting
 density of energy states is almost flat, i.e.
\begin{align}
\sum_{\vv x} w(E(\vv x)) \delta(E-E(\vv x)) &\approx \text{const}\;,
\end{align}
within a relevant energy window. In other words, the inverse of $w(E)$ is a rough approximation of the density of states $\rho(E)$.
This guarantees better mixing for systems with pronounced multi-modal structures.
In principle, a flat density can also be strived for other  parameters, e.g. magnetization.
The weight $w(E(\mathbf{x}))$ is not known in the beginning, otherwise we would already know $\rho(E)$ and a direct evaluation of $Z$ as sum over energies would be an easy task. 
It is generally iteratively determined by  repeated  simulations. 
More details can be found in \cite{Janke_1998,berg_introduction_1999,berg_multicanonical_2003}.
Apart from a yet unknown normalization constant $Z_{w}$, we define a new probability density function (PDF) for a configuration $\mathbf{ x}$ as 
$p^{(0)}(\mathbf{ x})=
w(E(\mathbf{ x}))/Z_{w}$.
The partition function can then be expressed as
\begin{align}
    Z_{\betaJ}  
      &= Z_{w}\sum_\mathbf{x}\; {e^{-\betaJ E( \mathbf{x})+g(E( \mathbf{x}))}\;
   p^{(0)}(E(\mathbf{x}) )  }\;.
\end{align} 
MUCA uses $p^{(0)}$ to guide the random walk and it yields a sample of configurations $\mathbf{x}$ of size $L$, say. We can then estimate the partition function by
\begin{align}
    Z_{\betaJ}
    &=\frac{Z_{w}}{L}
    \sum_{n=1}^{L}\;  e^{-\betaJ E( \mathbf{x}_{n})+g(E( \mathbf{x}_{n}))}
    =Z_{w}    \sum_{E}\;  e^{-\betaJ E+g(E)} \;h(E)\;,
\end{align} 
where $h(E)$ is the relative frequency of the occurrence of energy $E$ in the Markov chain.
Similarly  for the internal energy we obtain
\begin{align}
U^\text{MUCA}_{\betaJ}   &=
\frac{
\sum_{E}\; E\;e^{-\betaJ E+g(E)}\;h(E)
   }{
\sum_{E}\;e^{-\betaJ E+g(E)}\;h(E)
   }\;.
\end{align}
For the partition function we still need the normalization $Z_{w}$. It can be determined via
the exact limiting case $\betaJ =0$, or $\betaJ\to \infty$. To this end we consider
\begin{align}
\frac{Z_{\betaJ}}{Z_{\betaJ'}} 
&=
\frac{
\sum_{E}
e^{-\betaJ E + g(E)} p^{(0)}(E) 
}{
\sum_{E}
e^{-\betaJ ' E +g(E)} p^{(0)}(E) }\;.
\end{align}
For the two limiting cases we have according to  \eq{eq:limits}
  $Z_{\betaJ=0}=q^{N}$ and   $Z_{\betaJ\to \infty}=q$ 
and the  MUCA estimate yields
\begin{subequations}\label{eq:Z:MUCA}
\begin{align}
Z_{\betaJ}
&=
q^{N} \;\frac{\sum_{E}e^{-\betaJ  E+ g(E)} h(E) }{\sum_{E}e^{g(E)} h(E) }\;
&&(\text{\it high temperature limit}) \\
Z_{\betaJ}
&=q \;\frac{\sum_{E}e^{-\betaJ (E-E_{0}) + g(E)} h(E) }{e^{g(E_{0})} h(E_{0}) }\;
&&(\text{\it low temperature limit}) \;,
\end{align}
\end{subequations}
where $E_{0}$ is the lowest energy of the model.

\subsection{Nested sampling}\label{ssec:nesa}

The goal is again  the numerical evaluation of the partition function in \eq{eq:thermoint0}
Later on we will present adequate terms for likelihood and  prior in  case of the Potts model.
Skilling \cite{skilling_nested_2006} proposed to express $Z$ as Lebesgue integral
\begin{eqnarray} \label{eq:Z2}
  Z =  \int \: X(L)\, d L\;,
\end{eqnarray}
where the integral runs over the likelihood values $L$ and $X(L)$ stands for the prior mass
\begin{align}\label{eq:X:L}
X(L) &=\SumInt{x} \;\pi(\vv x) \;\theta\big( L(\vv x) > L\big)\;.
\end{align}
Equivalently, the partition function can then be expressed as
\begin{eqnarray} \label{eq:Z2b}
  Z =  \int_{0}^{1} \: dX \, {\cal L}(X)\;.
\end{eqnarray}
The derivation of the last three equations can be found in the appendix  \ref{app:NESA:Z}. 
Some minor modifications in the derivation  are due in the case of a discrete configuration space. They 
are outlined in section \ref{sec:NESA:details}.
Additional information can be found in \cite{von_der_linden_bayesian_2014}.
Finally, the integral in \Eq{eq:Z2} is approximated by a Riemann sum
\begin{eqnarray} \label{eq:riemann_s}
  Z =  \sum_{n=0}^{\infty} \: {\cal L}(X_{n}) \Delta X_{n} \quad , \quad
  \Delta X_{n} = X_{n}-X_{n+1} \;.
\end{eqnarray}
Because the likelihood is a monotonically decreasing function of the prior mass,
\eq{eq:riemann_s} represents a lower bound of the integral. Replacing ${\cal L}(X_{n})$ by
${\cal L}(X_{n+1})$ yields an upper bound.
It is needless to say that the computation of ${\cal L}(X)$ is as complicated as the original evaluation of $Z$, but due to Skilling \cite{skilling_nested_2006}, an algorithm can be constructed that avoids the actual knowledge of ${\cal L}(X)$. The pseudo code is given in \ref{algo_nest}.
  \begin{center}
 \begin{pseudocode}[ruled]{NESA algorithm}{\{\myhat{\lambda}_{n}\},n_\text{max}} \label{algo_nest}
  \textbf{input parameters: } K,k,\epsilon_{\lambda} \\
 \textbf{initialize } \myhat{\lambda}_{0}=0,\: n=0, \: \\
 \text{draw K configurations}\: \{x_{i}\}\: \text{ at random from } \pi(\mathbf{x}\CS \myhat{\lambda}_{0})\\
 \text{sort likelihood values $\lambda_{i}={\cal L}(x_i)$ in increasing order} \\
 \text{determine the $k^{\text{th}}$ smallest likelihood, denoted by }\myhat{\lambda}\\
\textbf{set } \myhat{\lambda}_{n=1}:= \myhat{\lambda} \\
 \WHILE \myhat{\lambda}_{n+1}-\myhat{\lambda}_{n} \: > \: \epsilon_{\lambda} \DO
 \BEGIN 
  n \GETS n+1\\ 
  \text{discard configurations with } \lambda_{i}\leq \myhat{\lambda}_{n}\\
    \text{replace them by new configurations as follows:}\\
 \text{\textbf{parallel}}\BEGIN
 \text{start thread $j =1,2,...,k$}\\
 \text{draw }\: {x_{n}^{j}}\: \text{ from } \pi(\mathbf{x}\CS \myhat{\lambda}_{n})\\
 \END\\
 \text{determine the $k^{\text{th}}$ smallest likelihood  $\myhat{\lambda}$
of all walkers}\\
\textbf{set } \myhat{\lambda}_{n+1}=\myhat{\lambda}
 \END\\
 \text{\textbf{set}  $n_\text{max}=n$}\\
 \RETURN{\{ \myhat{\lambda}_{n}\},n_\text{max}}
 \end{pseudocode}
 \end{center}
 %
In the initialization it is assumed that  likelihood values are not negative (hence $\myhat{\lambda}_{0}=0$), which will be the case for the Potts model.
During the NESA simulation $K$ configurations $\mathbf{x}$ (walkers) are treated simultaneously. In each step $k$ of the walkers, those with the smallest likelihood values, are replaced by new configurations, drawn from the prior subject to the constraint $L(\mathbf{\vv x})> \myhat{\lambda}_{n}$. The replacement of $k$ walkers is ideally suited for parallelization
(see \Se\ref{ssec:parallel_ns}). 
The nested sampling moves in configuration space ensure
that even well separated peaks of the likelihood function in configuration space are sampled correctly
(see \cite{von_der_linden_bayesian_2014}).
The crucial step for the nested sampling algorithm is to draw from
the constrained prior probability 
\begin{eqnarray} \label{eq:prior_res}
 \pi(\mathbf{x}\CS \myhat{\lambda}_{n-1}) = \frac{\pi(\mathbf{x})}{X(\myhat{\lambda}_{n-1})} \:\Theta (L(\mathbf{x})\:>\: \myhat{\lambda}_{n-1})\;,
\end{eqnarray}
which represents the normalized prior restricted to areas, where $L(\mathbf{x})$ exceeds the $\lambda$ threshold.
There exist various ways to draw random configurations from this prior.
In the approach employed here,  new walkers are determined by choosing $k$ of the remaining $K-k$ walkers at random, since they already represent a valid sample of the constraint prior, and to modify them by suitabel MC steps.
Given the likelihood minima $\myhat{\lambda}_{n}$, the Riemann sum in \Eq{eq:riemann_s} is  estimated by
\begin{eqnarray} \label{eq:ski_est}
 Z_\text{NESA} =  \sum_{n=1}^{\infty} \: \myhat{\lambda}_{n} \Delta {X}_{n}.
\end{eqnarray}
We assume that the likelihood has an upper limit
$\lambda_\text{max} = \max_{\vv x} L(\vv x)$, which will be the case for the Potts model,
and which will be reached at step $n=n_\text{max}$. Then we can stop 
the nested sampling run and proceed as follows
\begin{align}
Z_\text{NESA} &=\underbrace{
 \sum_{n=0}^{n_\text{max}-1} \myhat{\lambda}_{n}\Delta   X_{n}
}_{\color{blue} = Z_\text{NESA}^{1}}
+ \myhat{\lambda}_\text{max}\sum_{n=n_\text{max}}^{\infty} \Delta   X_{n}\;.
\end{align}
The second term yields
\begin{align}\label{eq:sum}
\sum_{n=n_\text{max}}^{\infty} \Delta   X_{n}
&=  X_{n_\text{max}}\;,
\end{align}
and we eventually have
\begin{align}\label{eq:Z:NESA}
Z_\text{NESA} &=Z_\text{NESA}^{1}
+ \myhat{\lambda}_\text{max} \;   X_{n_\text{max}}\;.
\end{align}
Based on the construction of the  threshold values $\myhat{\lambda}_{n}$ according to  the pseudo code \ref{algo_nest}, it has been shown by Skilling that the corresponding  prior masses have a simple and universal probability distribution\footnote{
Details are outlined in \cite{von_der_linden_bayesian_2014} and in appendix \ref {app:NESA:Z}
and \ref{sec:NESA:details}}.
First of all, the prior masses can be
expressed as
\begin{align}\label{eq:X:n}
X_{n} &=\prod_{\nu=1}^{n} \theta_{\nu}\;,
\end{align}
where the {\it shrinkage factors} $\theta_{\nu}$  are iid random variables. The PDF 
$p(\theta_{\nu})=p(\theta_{1})$ is the $k$-th order statistic 
of the uniform PDF and is therefore
a beta distribution,
\begin{align}\label{eq:beta:k}
p(\theta_{1}) &=  \frac{\big(\theta_{1}\big)^{K-k}\big(1-\theta_{1}\big)^{k-1}}{B(k,K-k+1)}\;.
\end{align}
For notational ease we have suppressed the explicit mention that $p(\theta_{1})$ depends on $k$ and $K$.
We can easily compute the mean of the prior masses $X_{n}$ 
\begin{align} \label{eq:avg:X:n}
\avg{X_{n}} &= \prod_{\nu}\avg{\theta_{\nu}}=\xi^{n}\;,\qquad \text{with  }\;\xi:=\frac{K}{K+k}\;.
\end{align} 
and  the increments 
\begin{align}
\Delta X_{n} &= X_{n}-X_{n+1}
=\prod_{\nu=1}^{n} \theta_{\nu} \bigg(1-\theta_{n+1}  \bigg)\;,
\intertext{or rather the logarithm of it}
\log(\Delta X_{n})&= \sum_{\nu=1}^{n}\log(\theta_{\nu})
+\log\big(1-\theta_{n+1}  \big)\;.
\end{align}
It will turn out that the logarithmic shrinkage factor $l_{\nu}=l_{1}=-\ln(\theta_{1})$ plays a crucial role. Its PDF follows directly from \eq{eq:beta:k} and is  simply an exponential for $k=1$
\begin{align}\label{eq:exp}
p(l_{1}) &= K e^{-K l_{1}}\;\qquad(\text{for } k=1)\;.
\end{align}
Mean and variance are (see e.g. \cite{henderson_parallelized_2014})   
\begin{subequations}\label{eq:fixedseq10}
\begin{eqnarray} 
\langle\: l_{1} \:\rangle &=&   \sum_{n=0}^{k-1}\frac{1}{K-n} \;
= \frac{\tilde k}{K}\\
\text{with}\quad \tilde k &:=& \sum_{n=0}^{k-1}\frac{1}{1-n/K}, \\
\langle (\Delta l_{1})^2 \rangle &=&  \sum_{n=0}^{k-1}\frac{1}{(K-n)^{2}} \;
=\frac{\tilde k' }{K^{2}}\\
\text{with}\quad \tilde k' &:=& \sum_{n=0}^{k-1}\frac{1}{(1-n/ K)^{2}} \;.
\end{eqnarray}
\end{subequations}
For $k\ll K$, which is in most circumstances a reasonable setting, we have $\tilde k =\tilde k' =k$. This is also valid for $k=1$, irrespective of $K$.

As argued by  Skilling, a special treatment is necessary, when  $Z$ becomes
extreme large. This is in particular the  case for partition functions, where $Z$ is exponential in the system size, as $\ln(Z)$ is an extensive quantity.
In this case the distribution of $Z$ will not be Gaussian any more, but rather that of $\ln(Z)$ \cite{skilling_nested_2006}. An example is given in \cite{von_der_linden_bayesian_2014}.
We will therefore  compute the probability for $\ln(Z)$, given the set of threshold values $ \vvmyhat{\lambda}=\{\myhat{\lambda}_{n}\}$.
To this end we marginalize over the set of shrinkage factors $\vv \theta =\{\theta_{n}\}$.
\begin{align}
p(\ln(Z)\CS  \vvmyhat{\lambda},{\cal I}) &= \int d \vv \theta \;p(\ln(Z)\CS  \vvmyhat{\lambda},\vv \theta,{\cal I}) \;p(\vv \theta\CS  {\cal I}) \notag \\
&= \int d \vv \theta \;\delta\bigg[\ln(Z)- \ln\big(  {\cal Z}(\vvmyhat{\lambda},\vv \theta)\big)\bigg]\;p(\vv \theta\CS  {\cal I}) \;.
\end{align}

where $p(\vv \theta\CS \BK) = \prod_{\nu} p(\theta_{\nu})$ and $d\vv \theta = \prod_{\nu} \theta_{\nu}$.
We can now  easily evaluate this expression numerically by drawing $N_{\text{pr}}$ realizations $\vv \theta^{(m)}$, of the set of shrinkage factors with 
$m=1,\ldots,N_{\text{pr}}$,
according to   \eq{eq:beta:k}
and estimate the lowest moments of $\ln(Z)$ via
\begin{align} \label{eq:moments:Z}
\avg{[\ln(Z)]^{\gamma}} &= \frac{1}{N_{\text{pr}}}
\sum_{m=1}^{N_{\text{pr}}} \left\{ \ln \bigg({\cal Z}(\vvmyhat{\lambda},\vv \theta^{(m)})\bigg)\right\}^{\gamma}.
\end{align}
The results for $\gamma=1,2$ allow to estimate mean and variance of $\ln(Z)$. It should be stressed again that all operations need to be performed in logarithmic representation. This shall be illustrated
for  the sum of any terms $a_{n}$. We first compute the logarithm $\alpha_{n}:=\ln(a_{n})$ and the maximum
$\alpha_\text{max}:=\max_{n} \alpha_{n}$.
Then
\begin{align}
\ln\bigg(\sum_{n} a_{n} \bigg)&= \alpha_\text{max} + \ln\bigg(\sum_{n} \exp\big( \alpha_{n}-\alpha_\text{max} \big)\bigg)\;.
\end{align}

In summary, we have presented, how the partition function can be evaluated by nested sampling in terms of a sample of configurations $\vv x_{n}$, with monotonically increasing  likelihood values. There are two sources of uncertainty, the discretization error due to the approximation of the integral by a sum (see \eq{eq:riemann_s}) and a statistical uncertainty that stems from the scatter of the likelihood values $\lambda^{*}_{n}$ about the unknown mean $\langle \lambda^{*}_{n}\rangle$. The discretization error decreases with increasing
number of walkers $K$ and the statistical uncertainty can be estimated by the PDF of the prior masses, which is analytically known. More details will be discussed in the frame of the Potts model.

\section{Application to the Potts model}\label{ssec:nsbasic}

\newcommand{\bino}[2]{\begin{pmatrix}#1\\#2\end{pmatrix}}

In \cite{murray_nested_2005} the application of the NESA algorithm to the Potts model in the representation of
Fortuin and Kasteleyn (FK) is introduced, where the spin variables are replaced by bond variables $b_{ij}$ defined between each pair of neighbouring sites $i$ and $j$.  A bond variable $b_{ij}$  is either active (1) or inactive (0). We denote the entire bond configuration on the lattice by $\mathbf{b}$. 
For a graphical representation each active bond $b_{ij}$ is represented by a line connecting the sites $i$ and $j$. 
The set of sites, connected by lines forms a cluster. An isolated site, to which no line is attached, also qualifies as cluster. In the FK representation two properties of a bond configuration $\mathbf{b}$ are of central importance, the number of active bonds 
\begin{align} \label{eq:activebonds}
D(\mathbf{b}) = \sum_{\langle i,j\rangle } b_{ij} 
\end{align} 
and the number of clusters $C(\mathbf{b})$ formed by the set of active bonds $\mathbf{b}$.   
In  the FK model  the distribution function for the bond variables $\mathbf{b}$ reads
\begin{eqnarray}
 P(\mathbf{b}\CS \betaJ) &=& \frac{e^{-\betaJ \Npair}}{Z_{\text{P}}} e^{\kappa\:D(\mathbf{b})}\: 
  q^{C(\mathbf{b})} 
\label{eq:nspotts}\;,
 \end{eqnarray}
where 
$\Npair$ is the number of pairs in the lattice, which is given by
$\Npair=2N$ for the 2d square lattice and $\kappa = \ln(e^{\betaJ}-1)$. 
The probability that nearest neighbours with equal spin value form an active bond is defined as
\begin{align} \label{eq:p:active:bond}
p_\text{b}=1-e^{-\betaJ}\;.
\end{align} 
The partition function $Z_{\text{P}}$ in the bond representation is equivalent to the spin-representation and it  reads
\begin{eqnarray}  \label{eq:pottsZdef}
 Z_{\text{P}} 
  = Z_{\pi}\:e^{-\betaJ\Npair}\:  Z_{\text{NESA}}  \label{eq:Zns:a} \quad, \quad
  Z_{\text{NESA}} :=\sum_{\mathbf{b}} L(\mathbf{b}) \: \pi (\mathbf{b})\;.
\end{eqnarray}
The likelihood function and the prior probability are defined as
\begin{eqnarray}\label{eq:Liklins}
L(\mathbf{b}) = e^{\kappa\:D(\mathbf{b})} \quad,\quad
\pi (\mathbf{b})=\frac{q^{C(\mathbf{b})}}{Z_{\pi }}  \;.\label{eq:priorix}
\end{eqnarray}
In order to have  a normalized prior $\pi (\mathbf{b})$ we 
had to introduce 
$Z_{\pi}= \sum_{\mathbf{b}}  q^{C(\mathbf{b})}$ as prefactor in \eq{eq:Zns:a}.
It is essential that the unknown prior normalization $Z_{\pi}$ is not a function of $\betaJ$.
To determine $Z_{\pi }$ we can use one of the two limit  cases $\betaJ =0$ or $\betaJ\to\infty$, for which the partition function is given in \eq{eq:limits}.
First we note that $\betaJ=\betaJ^{*}=\ln(2)$ splits the temperature into two regimes, since 
for $\betaJ> \betaJ^{*}$   ($\betaJ\le \betaJ^{*}$) we have $\kappa >0$ ($\kappa <0$). Since nested sampling requires monotonically increasing likelihood values, $\betaJ> \betaJ^{*}$   ($\betaJ\le  \betaJ^{*}$) corresponds to increasing (decreasing) $D(\mathbf b)$. We therefore have to perform
separate NESA runs for these two temperature  regimes. We are, however, only interested in
$\betaJ > \ln(2)$ as it includes the phase transition and the low temperature regime.
In this temperature regime the likelihood constraint has the form $D(\mathbf{b})>D^*$.
It also includes the limit $\betaJ\to \infty$. For this limit 
the exact value of $Z_{\text{P}}$ is given in \eq{eq:limits}, and we rewrite
 \eq{eq:Zns:a} as
\begin{equation}\label{eq:Zns:c}
\begin{split}
\ln\big(Z_{\pi }\big) 
&=\lim_{\betaJ \to \infty} \bigg(\Npair \betaJ +
\ln\big[Z_\text{P}(\betaJ)\big] - \ln\big[Z_\text{NESA}(\betaJ)\big]\bigg)\\
&= \ln(q) + 
\lim_{\betaJ \to \infty}
\bigg( \Npair\betaJ- \ln\big[Z_\text{NESA}(\betaJ)\big]\bigg)
\;.
\end{split}
\end{equation} 
Hence, the prior normalization can  be determined if we can determine  $Z_\text{NESA}(\betaJ\to \infty)$ reliable from NESA, which is indeed easily the case, as we shall see later (\Se\ref{sec:stoch:eval:Z}).

\subsection{Technical details of the nested sampling algorithm}\label{ssec:NESAalgorithm}

In this section we want to discuss some technical details of the NESA algorithm for the Potts model closely related to the ideas outlined in \cite{murray_nested_2005}. First of all, we need a Markov Chain MC algorithm to sample from the constrained prior. Interestingly, for $\kappa=0$, i.e. $ \betaJ= \betaJ^{*}$ the distribution function of the Potts model (see \eq{eq:nspotts}) coincides with the prior. So we can simply apply Swendsen-Wang (SW) for that temperature, which corresponds to  $p_{b}=1/2$. Also the likelihood constraint can easily be incorporated, as we shall discuss next.

Initially we have $\lambda_{0}=0$ and $D_0=0$, i.e. there is no likelihood constraint, and we 
simply sample from the prior, by applying SW with $p_{b}=1/2$. 
Then the next steps are as follows:

\begin{enumerate}
  \item Identify the clusters.
  \item For each cluster draw a random spin value $\in\{1,\ldots,q\}$ that is assigned to all spins of the cluster.
  \item Identify the list ${\cal L}$ of nearest neighbour pairs 
with equal spin values. Only elements in ${\cal L}$ can become active bonds. Let $D_{c}$ be the number of bond candidates in ${\cal L}$.
  \item For each element in ${\cal L}$ assign an active bond with probability  $p_b=1/2$.
\end{enumerate}
The result yields a new bond configuration $\mathbf{b}'$ drawn according to the prior probability.
For the cluster identification we employ the Weighted Quick Union Find with Path Compression
(WQUPC) \cite{Algorithms_2011} algorithm of Robert Tarjan, which requires an operation count of $N\ln^{\star}(N) $, where $\ln^{\star}(N)$ is defined as the smallest $n$ with 
\begin{align}
\underbrace{
\ln(\ln(\ldots\ln(
}_{\color{blue} = n}N) < 1\;.
\end{align}
For the $N=512\times 512$ system $\ln^{\star}(N) = 3$ and even 
for a $N=10^{6}\times 10^{6}$ system $\ln^{\star}(N)$ is merely 5.

Before we can implement the likelihood constraint we have to get rid of likelihood degeneracies. There are many bond configurations with the same number of active bonds. As outlined in \cite{von_der_linden_bayesian_2014} the degeneracy can be lifted by augmenting the phase space by a single additional variable, $x$ say. 
The walkers now consist of the bond configuration $\mathbf{b}$ and the value $x$.
We introduce a modified likelihood defined as
\begin{align}\label{eq:likelihood:aug}
L(\mathbf{b},x) &:= e^{\kappa\big(D(\mathbf{b}) + x \;0^+  \big)}\;,
\end{align} 
where $0^+$ is an  infinitesimal positive real number. For 
 the joint distribution function we use
\begin{align}
P(\mathbf{b},x) := P(\mathbf{b}) \;p(x) \quad
\text{with} \quad p(x) :=p_{u}(x\CS x_{0}=0)\;,
\end{align} 
where $p_{u}(x\CS x_{0})$ is the PDF of a uniform random variable
from the interval $(x_{0},1]$, i.e.
\begin{align} \label{eq:uniform:pdf}
p_{u}(x\CS x_{0}) &:=
\frac{1}{1-x_{0}} \Theta(x_{0} < x \le 1)\;.
\end{align} 
The additional variable $x$ in the augmented likelihood 
in \eq{eq:likelihood:aug} lifts the degeneracy and has otherwise no impact on the  likelihood values.
The likelihood constraint in the augmented phase space reads
\begin{align} \label{eq:l:constraint:new}
D(\mathbf{b}') + 0^{+} x' > D^* + 0^{+} x^*.
\end{align} 
Obviously, only in the case of degeneracy the auxiliary variable $x$ comes into play.
The implementation of this constraint is now in principle an easy task. Given the threshold pair $(D^*,x^*)$ we  draw from the prior at random a new walker configuration $(\mathbf{b'},x')$. If it fulfills the likelihood constraint (\eq{eq:l:constraint:new}) the new configuration is accepted and it is rejected otherwise. 
The rejection step can, however, become very time consuming.
Therefore it is advisable  to avoid the rejection steps.
According to the above rules, bonds are independently activated with probability $p_b=1/2$. The number of active bonds, therefore,  follows
a binomial distribution ${\cal B}(D'\CS D_{c},p=\frac{1}{2})$. Due to the likelihood constraint we need the truncated binomial
\begin{align} \label{eq:runc:bino}
  \tilde P(D'\CS D^{*},D_{c}):=\frac{\Theta(D'\ge D^{*})}{Z_{B}(D^{*},D_{c})} {\cal B}(D'\CS D_{c},p=\frac{1}{2}) , 
\end{align} 
  where $Z_{B}(D^{*},D_{c})$ is the corresponding normalization.
  The cumulative distribution function is defined as $P_{D'>D^*}:= \tilde P(D'>D^*\CS D^*,D_{c})$.
  Drawing the new number of active bonds $D'$ from $P(D'\CS D^{*},D_{c})$ includes the case $D'=D^*$
  where the auxiliary variables $x'$ and $x^*$ have to ensure that the likelihood values
  in augmented phase space $D+0^{+}x$ increase monotonically.
  By the elementary rules of probability theory we find easily that the probability, that the next accepted step in the brute-force approach corresponds to  $D'> D^*$, is given by
\begin{align}
\tilde P_{D'> D^*}=\frac{P_{D'>D^*}}{1-(1-P_{D'>D^*}) P_{x'<x^*}}.
\end{align} 
with $P_{x'<x^*}:=P(x'<x^*) = x^*$.
In opposite case the next accepted step comes from $D'= D^*$ with the auxiliary variables $x'>x^*$.
Now, we can modify step 4 of the SW algorithm to incorporate the likelihood constraint in an rejection-less way:
\begin{enumerate}
  \item[4a.] Draw a random number $r$ from  $p_{u}(x\CS x_{0}=0)$.
  \item[4b.] If $r< \tilde P_{D'>D^*}\; (\text{i.e. }  D'>D^*)$
  \begin{itemize}
    \item determine at random the number of active bonds $D'$ according to  $\tilde P(D'\CS D^*+1,D_{c})$
    \item and draw at random $x'$ from $p_{u}(x\CS x_{0}=0)$.
  \end{itemize}
  \item[4c.] If $r\ge \tilde P_{D'>D^*}\; (\text{i.e. } D'=D^*)$
    \begin{itemize}
       \item draw at random  $x'$ from $p_{u}(x\CS x_{0}=x^*)$
       \item and set $D' = D^*$.
     \end{itemize}
    \item[4d.] Activate at random $D'$ bonds from the list ${\cal L}$, resulting in the new bond configuration $\mathbf{b}'$.
    \item[4e.]  The new walker configuration is $(\mathbf{b}',x')$.
\end{enumerate}

Finally it should be stressed that a single NESA run 
suffices to compute  the partition function for all temperatures 
$\betaJ> \ln(2)$. This is easily achieved by storing
the number of active bonds $\{\myhat{D}_{n}\}$ instead of the corresponding likelihood minima $\{\lambda^{*}_{n}\}$, introduced before. Based on \eq{eq:Liklins}
we can determine the likelihood values 
$\myhat \lambda_{n} = e^{\kappa \myhat{D}_{n}}$ for all temperatures and compute the partition function according 
to \eq{eq:ski_est}.

In the case of the Potts model, nested sampling yields the sequence of active bonds $\myhat{D}_{n}$,
which correspond to the prior masses $\myhat X_{n}$. The meaning of the latter corresponds in this case to the  probability $P(D \geq D^{*})$, or rather  the complementary cumulative distribution function, from which we obtain easily the probability $P(D)$. By construction, the probability for the number $D$ of active bonds is closely related to the probability for the number $N_\text{eq}$ of equal spins on neighboring sites, and that in turn is trivially related to the  density of states $\rho(E)$, discussed in the context of the multi-canonical sampling. 
Nested sampling therefore offers the possibility to compute $\rho(E)$ as well. 
But we discuss more
direct ways to compute physical quantities of interest in nested sampling in the next sections.

\subsection{Autocorrelation times}\label{ssec:autocorrtime}

A key element of nested sampling, and as a matter of fact the only place where autocorrelation could play a role, is the generation of  configurations according to the constraint prior.
This is achieved, as described in \Se\ref{ssec:nesa}, by starting from a configuration that fulfills already the constraint and then
$N_{c}$ repeated SW updates are performed that also fulfill the likelihood constraint.
Such configurations may exhibit autocorrelations, which
shall be analyzed in this section.
But it should be remembered right at the beginning, that SW just plays an auxiliary role and is
only required at an effective inverse  temperature  $\betaJ = \ln(2)$, far away from the critical point. 
Given a time series of bond configurations, $\mathbf{b}^{(n)}$ for $n=1,2,\ldots,N_{c}$, we compute the autocorrelation in the number of active bonds 
$B_{n}:=D(\mathbf{b}^{(n)})$, i.e.
\begin{align}\label{eq:ac}
\rho_{m} 
&:= \frac{1}{M} \sum_{n=1}^{M} \Delta B_{n+m}\;\Delta B_{n}\;,
\end{align} 
with $M= N_{c}-m_\text{max}$, and $m_\text{max}$ being the maximum lag $m$ 
for which the autocorrelation is computed. We proceed as follows:
a single NESA run is started from a given initial configuration $\mathbf{b}^{(0)}$ 
with the corresponding likelihood $\lambda^*:=L(\mathbf{b}^{(0)})$ and 
a fixed number $N_{c}$ of SW updates $\mathbf{b}^{(n)}\to \mathbf{b}^{(n+1)}$  restricted to $L(\mathbf{b}^{(n+1)})\ge \lambda^*$ is generated. The threshold $\lambda^*$ is not modified during these steps.
The sequence of bond configurations $\mathbf{b}^{(n)}$  are now used to compute the corresponding sequence $B_{n}=D(\mathbf{b}^{(n)})$ of active bonds 
and the autocorrelation defined in \eq{eq:ac}.
Now, for one and the same initial configuration we repeat the elementary NESA run 
$L$ times (they only differ in the random numbers) and average the individual autocorrelation functions $\rho_{m}$ resulting in $\overline \rho_{m}$.
Next we determine  the integrated autocorrelation time $\tau_\text{int}$ by the  following procedure.
The average autocorrelation function $\overline\rho_m(t)$ is  cut off at $m^*$, where either an increase or a negative value in $\overline\rho_m(t)$ occurs.  This is necessary to get rid of the statistical noise in the data. Then we append a single-exponential tail to 
$\overline\rho_m$ for $m>m^*$. The parameters are determined from the second half of the truncated data ($m^*/2< m \le m^*$). Finally the integrated correlation time $\tau_{int}$ is computed by summing $\overline\rho_m(t)$ up to the cut-off value and then adding the contribution of the exponential tail, which can be expressed analytically by a geometric sum \cite{baillie_comparison_1991}.

\begin{figure}[t!]
\centering
\includegraphics[width=0.5\textwidth]{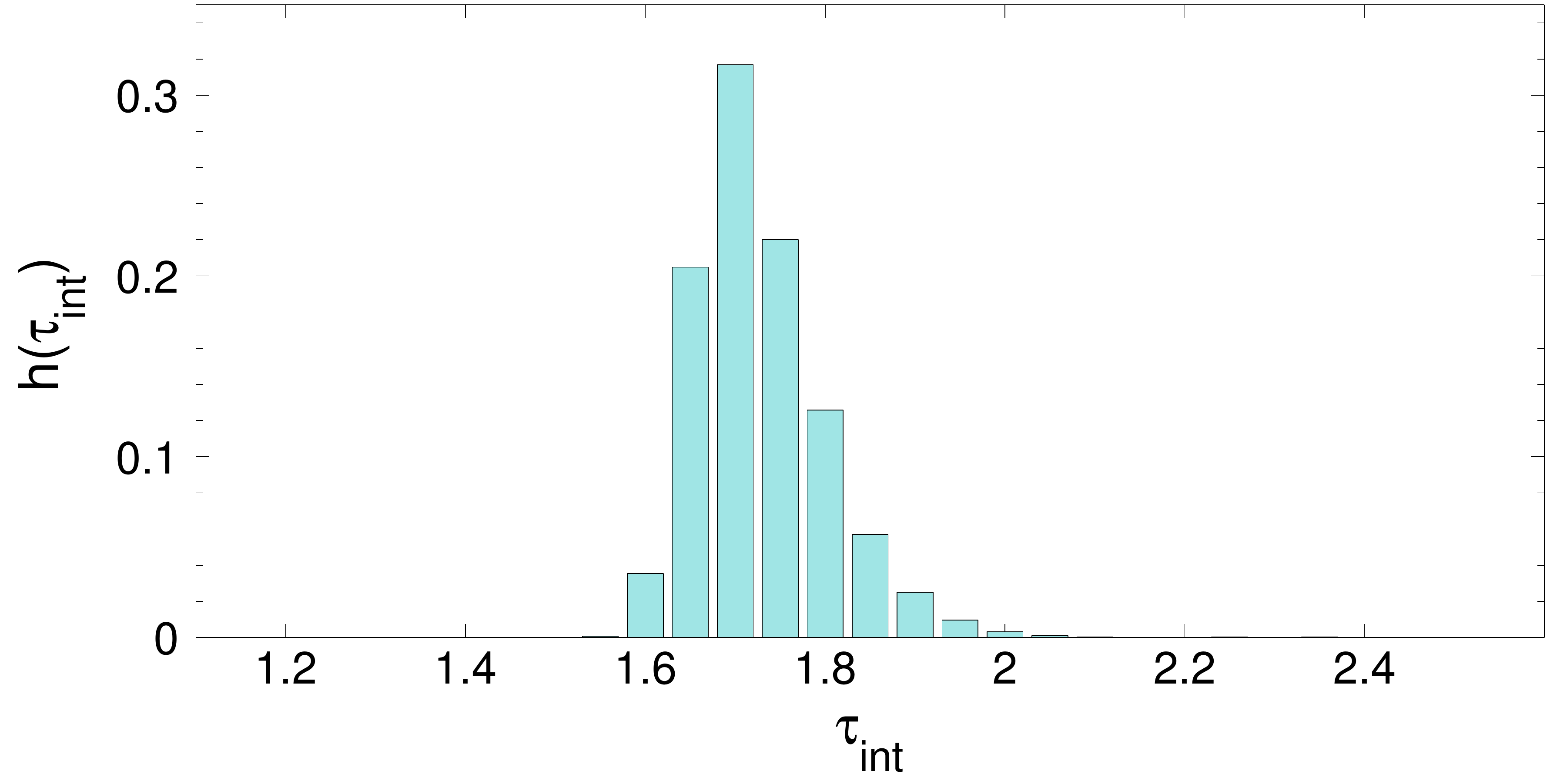}
\captionsetup{justification=raggedright,singlelinecheck=false}
\caption{Histogram of the integrated correlation times $\overline \tau_{int}$ corresponding to the autocorrelation function in \eq{eq:ac} for a $16\times16$ Potts model with $q=10$.\label{fig:hist_corrtime}} 
\end{figure}

Finally we analyse the impact of the initial configuration $\mathbf{b}^{(0)}$. To this end we perform $L$ elementary NESA runs of length $2 N$, all starting from the initial bond configuration $\mathbf{b}_{0}$ with fixed threshold $\lambda^*$ (as before) 
    and from all $L$ final configurations we  pick the one with the least likelihood and use it as new initial configuration $\mathbf{b}^{(0)}$. 
For $M$ different initial configurations thus determined  we compute individually the integrated autocorrelation times. The actual numbers used are $N=L=100$ and $M=1000$.
The histogram of the  integrated correlation times $\overline \tau_{int}$ 
of a $16\times 16$ system for $q=10$   is shown in figure \ref{fig:hist_corrtime}. 
Obviously, the for the $16\times 16$ system, the mean of the distribution is roughly $1.75$ with a small standard deviation.
For other system sizes and $q$-values we make essentially the same observation. 
So we can make the important conclusion that there is essentially no autocorrelation in that part 
of NESA, where we sample according to the constraint prior.
That may sound surprising since the SW algorithm is known to mix slowly for the Potts model with $q > 4$ \cite{gore_swendsen_1999}, but this is
only the case near the transition temperature $\betaJ$.
Moreover, in NESA the elements of the Markov chain, $\lambda_{n}^{*}$, do not really follow 
from the  SW trajectory of a single walker, but there are jumps between walkers,
that additionally lead to de-correlation.

\subsection{Performance of nested sampling}\label{ssec:z_and_ddz}
 
We are interested in the dependence of the partition function on the inverse temperature $\beta$, as it 
 provides the entire thermodynamic information of the system. As emphasized before, the dependence of $\ln(Z(\betaJ))$ on $\betaJ$ (for $\betaJ>\betaJ^{*}$, i.e. $\kappa>0$) can be obtained from a single NESA run.
\begin{figure}[t!]
\centering
\includegraphics[width=0.5\textwidth]{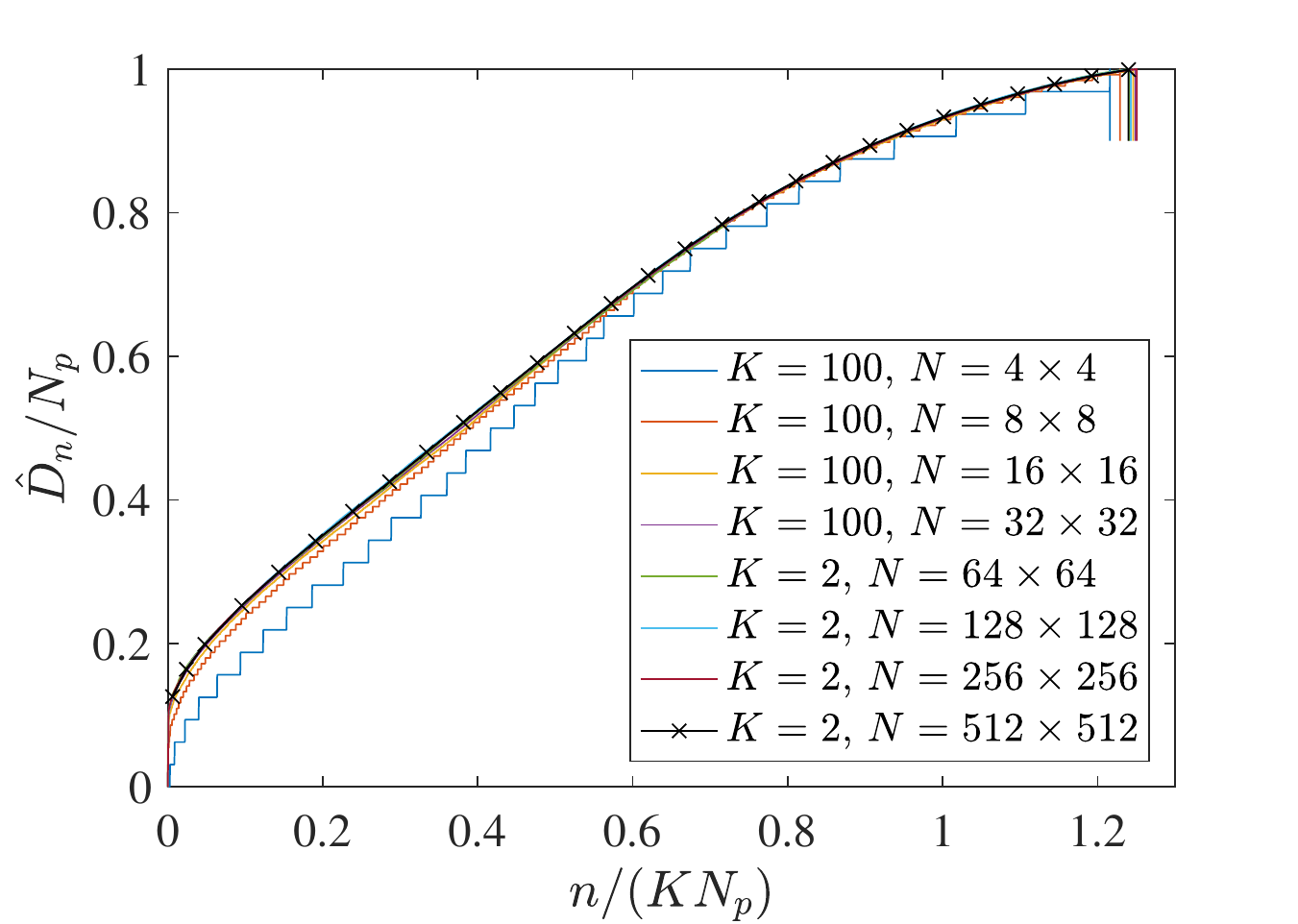}
\captionsetup{justification=raggedright,singlelinecheck=false}
\caption{$\myhat{D}_{n}$ for the $q=10$  Potts model with  $K=2$ and $K=100$ and $k=1$ for
various system sizes. The short vertical bars at the end mark $n_\text{max}$.\label{fig:D_n}} 
\end{figure}
Instead of generating the increasing sequence of likelihood values, we can equally well generate the corresponding increasing sequence 
of the active bonds numbers $D_{n}$. The upper limit of the number of active bonds $D_\text{max}$ is equal to the number of pairs $\Npair$. In \fig{fig:D_n} we depict the fraction  $\myhat{D}_{n}/\Npair$ versus $n/(K \Npair)$ for various system sizes
and walkers. All plots are obtained for $k=1$, i.e. in each NESA step, one walker was updated.
Obviously, with increasing system size the results rapidly converge  towards  a universal curve. An important finding is that  the number of NESA steps $n_\text{max}$ it takes to reach $D_\text{max}$ is proportional to the number of pairs and the number of walkers. 
A separate study for $k\in\{1,\ldots,10\}$ and  $q\in\{2,\ldots,10\}$ yields the following scaling behavior
\begin{align}
n_\text{max} &= (a + b q +  c q^2) \frac{K \Npair }{k}\;,
\end{align} 
with $a= 0.606, b= 0.096, c = -0.003$ for a $8\times 8$ system
and $a= 0.599, b= 0.089, c= -0.002$ for a $16\times 16$ system.
We find that the $q$-dependent prefactor is nearly independent of the 
system size and it is a very smooth function in $q$. There is no distinction in the behaviour for systems that have a first  or second order phase transition.
The key message so far is, however, that NESA needs  $O(1) \frac{K \Npair}{k}$ steps.
An rigorous analytic proof of this finding is given in appendix \ref{app:n:max}.

\subsection{Properties of the threshold values $\myhat{D}_{n}$}

\begin{center}
\begin{figure*}
{\includegraphics[width=0.33\textwidth]{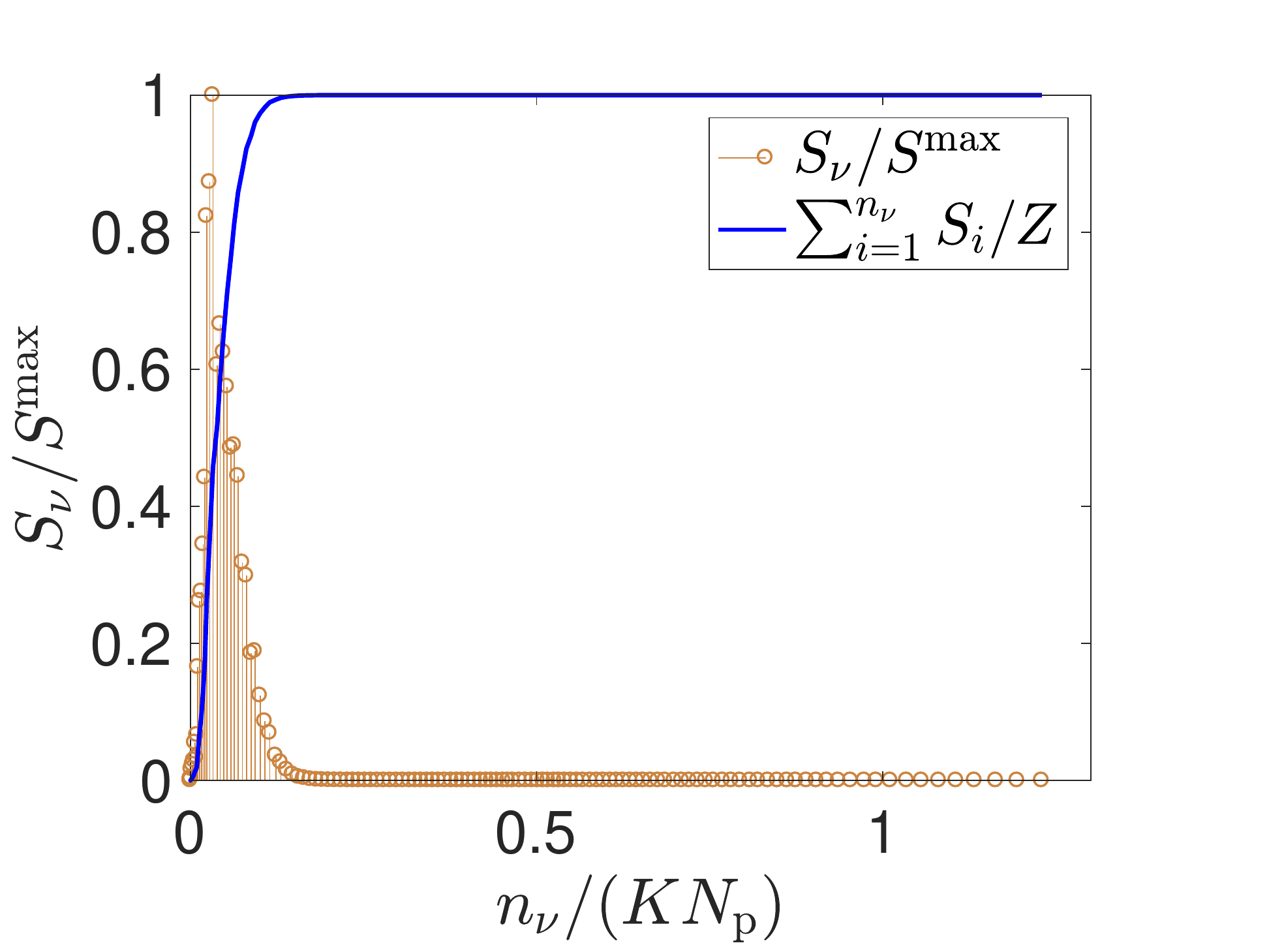}} 
\hspace{-1mm}
{\includegraphics[width=0.33\textwidth]{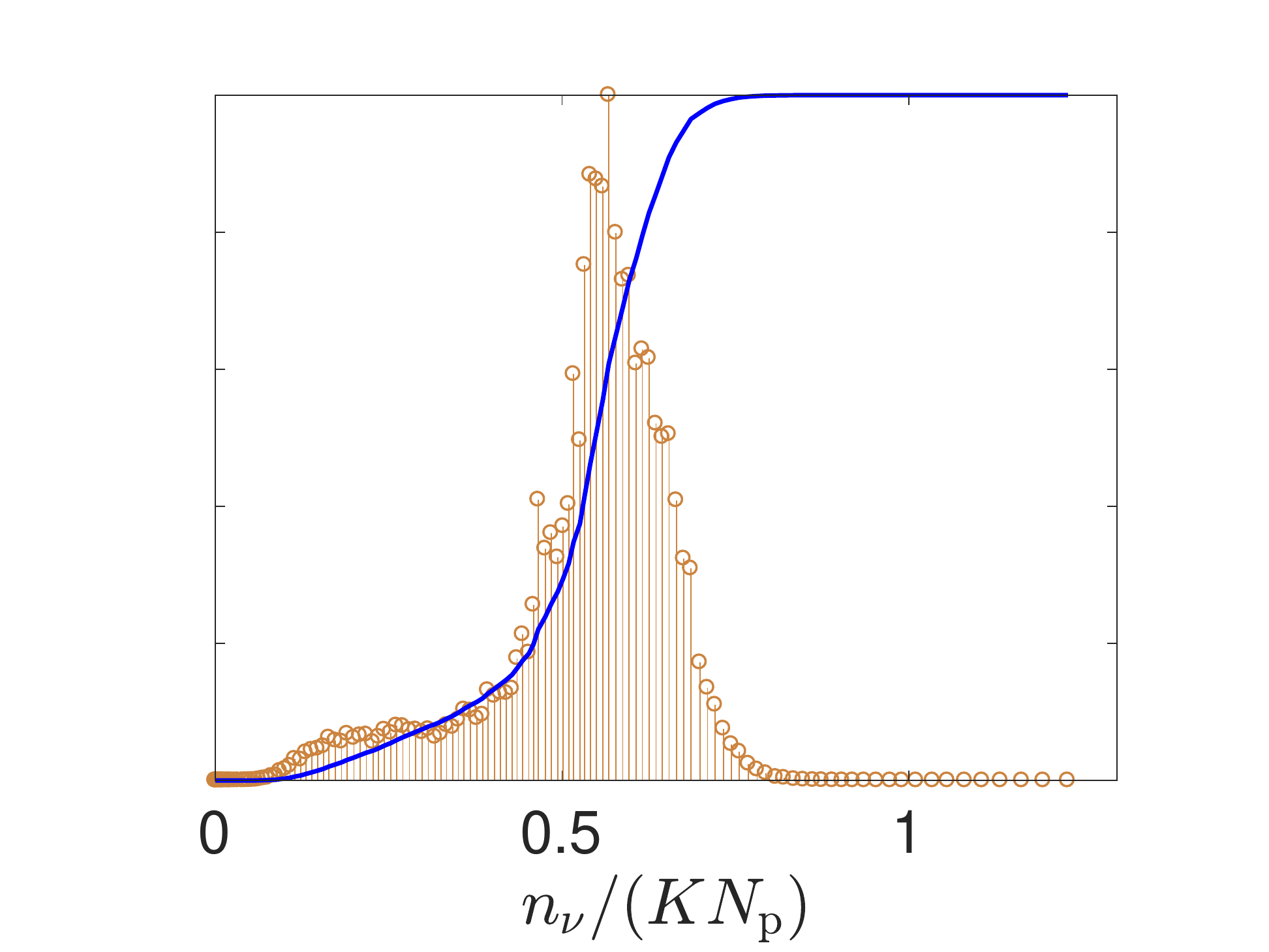}}
\hspace{-2mm}
{	\includegraphics[width=0.33\textwidth]{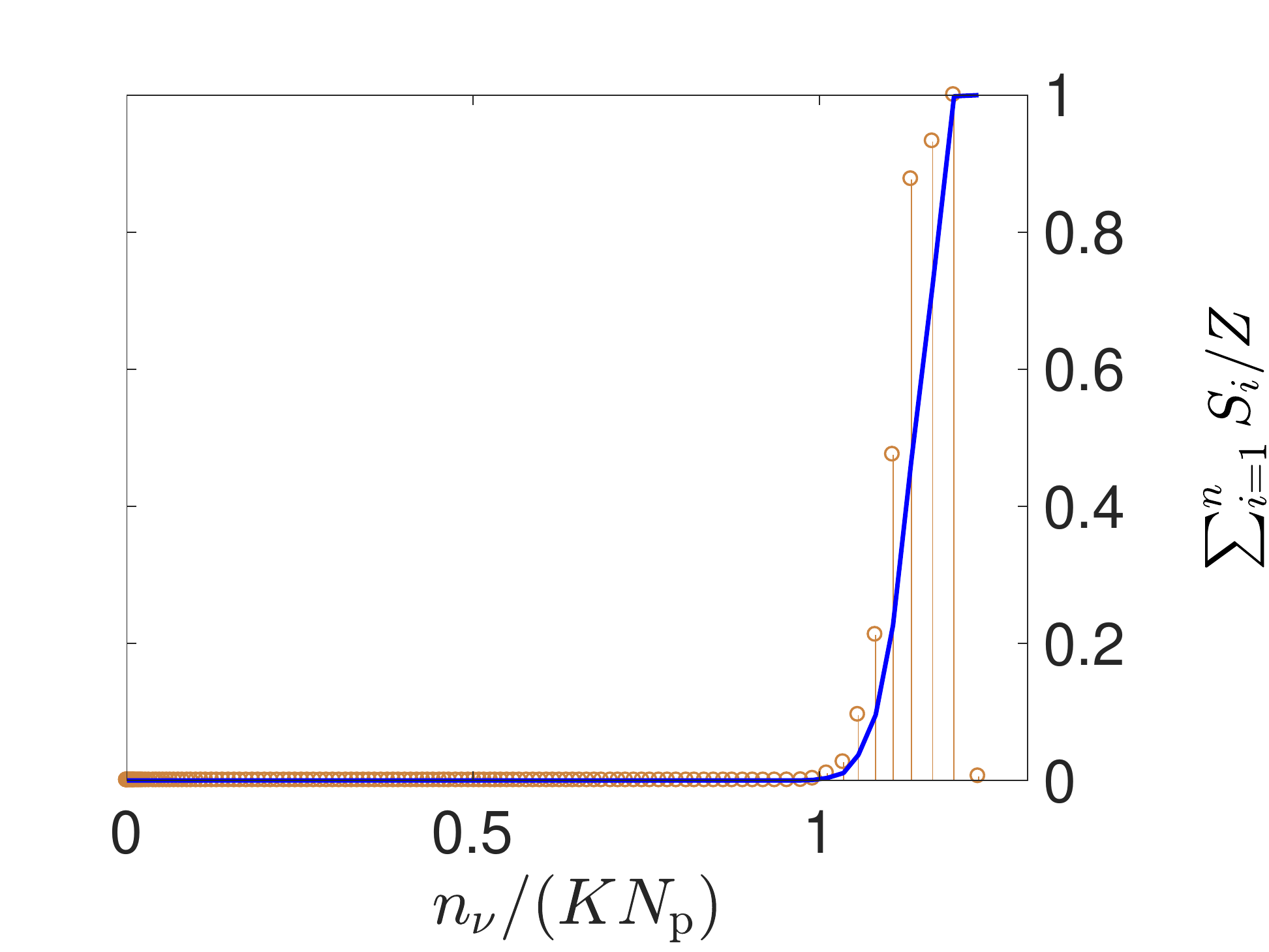}}
\caption{Summands $S_{\nu}$ of \eq{eq:aux11} and the corresponding cumulative sums for a  $q=10$ Potts model on a $8\times8$ lattice at different temperatures (from left to right): $\betaJ= 1$,  $\betaJ_c = \betaJ_{c}$,  $\betaJ= 4$. With increasing $\betaJ$  the main contribution to the sum shifts from the far left to the far right. }
\label{fig:lofx2}
\end{figure*}
\end{center}

Next we shall study the properties  of the threshold values $\myhat{D}_{n}$ for the case $k=1$. The results for various system sizes and walker numbers are shown in
\fig{fig:D_n}.
We see that $\myhat{D}_{n}$ represents a stairway with
 steps that have an average height of $\approx 1$ and average width of $\approx K$. 
We denote the position, at which  
 the $\nu$-th step begins, by   $n_{\nu}$ and the corresponding height by $\myhat{D}_{n_{\nu}}$.
Based on  \eq{eq:avg:X:n} we find $\avg{\Delta X_{n}} = \xi^{n} (1-\xi)$ and the mean partition function can then be written as
\begin{align} \label{eq:aux11}
\avg{Z_\text{NESA}(\betaJ)} &= \sum_{\nu} S_{\nu}\;,\qquad\text{with }\;\;
S_{\nu}:=e^{\kappa \myhat{D}_{n_{\nu}}} \;(\xi^{n_{\nu}}-\xi^{n_{\nu+1}})
\end{align} 
The summand $S_{\nu}$ represents the contribution of the $\nu$-th step  to the partition function, which clearly depends on temperature. In \fig{fig:lofx2} the normalized summands $S_{\nu}/S^\text{max}$ ($S^\text{max}=\max_{\mu}S_{\mu}$) are plotted as function of the
 step position $n_{\nu}$ scaled by $K \Npair$ for three inverse temperatures, below, at, and above the critical temperature $\betaJ_{c}$.
 For $\betaJ=1$ one observes that only a small fraction of the stairway  
in \fig{fig:D_n}  is sufficient for a converged result. For $\betaJ_c$ already a significant fraction is required and for  $\betaJ=4$ the last summands clearly dominate the partition function. 
Hence for  low temperatures (large $\betaJ$) an accurate estimate
for the partition function is only possible, if the NESA algorithm reaches the maximal number of active bonds $\myhat{D}_{n}= \Npair$. 
Therefore the CPU-time scales like $N^2$, because the number of NESA steps required to reach 
the maximum likelihood value is  $n_\text{max}$, which is proportional to $N$, and each NESA step involves one Swendsen-Wang update, which also scales like $N$, if WQUPC is used for the cluster identification. 

\subsection{Partition function \label{sec:stoch:eval:Z}}

We are now in the position to determine the yet unknown   prior normalization $Z_{\pi }$ within the same  NESA run. According to \eq{eq:Zns:c} we need $Z_\text{NESA}(\betaJ\to\infty)$.
For $\betaJ \to \infty$ we have $\kappa=\betaJ$ and \eq{eq:riemann_s} yields 
\begin{align}
Z_\text{NESA}(\betaJ) &= \sum_{n=0}^{\infty} e^{\betaJ \myhat{D}_{n}} \Delta X_{n}\;,
\end{align} 
from which for $\betaJ\to \infty$ only the terms with the maximal value of $\myhat{D}_{n}$ (i.e. $\myhat{D}_{n}=N_{p}$) determines the result, which is the case for $n\ge n_\text{max}$. We therefore have
\begin{align}
Z_\text{NESA}(\betaJ ) &\underset{\betaJ\to \infty}{\longrightarrow} e^{\betaJ N_{p}} \;\sum_{n=n_\text{max}}^{\infty} \Delta X_{n}\;.
\end{align} 
The latter sum yields
\begin{align}\label{eq:sum}
\sum_{n=n_\text{max}}^{\infty} \Delta X_{n}&=
\lim_{L\to \infty} \sum_{n=n_\text{max}}^{L} \Delta X_{n}
=X_{n_\text{max}}\;,
\end{align}
and we find for \eq{eq:Zns:c}
\begin{align}\label{eq:ln:Z:pi}
\ln\big(Z_{\pi }\big) &= \ln q - \ln X_{n_\text{max}}\;.
\end{align} 
Inserting this result in \eq{eq:pottsZdef} yields
\begin{align}\label{eq:ln:Z:new}
\ln\big(Z_{\text{P}}(\betaJ)\big) &= \ln q - \ln X_{n_\text{max}} -\betaJ N_{p} + \ln\big(Z_\text{NESA}(\betaJ)  \big)\;.
\end{align} 
In particular for $\betaJ = \betaJ^{*}$ (i.e. $\kappa=0$) we have
\begin{align}\label{eq:lnZ:Xmax}
\ln\big(Z_{\text{P}}(\betaJ^{*})\big) &= \ln q - \ln X_{n_\text{max}} -\betaJ^{*} N_{p}\;.
\end{align} 
Since $\ln(Z_{\text{P}})$ is an extensive quantity in the thermodynamic limit, for all temperatures, for large $N$ we therefore find 
\begin{align}\label{eq:X:max}
\ln X_{n_\text{max}}\propto N\;.
\end{align}
This expression is used in the numerical evaluation of $\ln(Z_{\text{P}}(\betaJ))$. 
$\ln (X_{n_\text{max}})$ turns out to play a crucial role, also in the following considerations. 
Its properties are studied analytically in appendix \ref{app:X:max}. 
For an initial qualitative discussion
we use  \eq{app:avg:X:max} and \eq{eq:fixedseq10} to compute the mean of  $\ln(X_{n_\text{max}})$, which is
$\avg{\ln(X_{n_\text{max}})}= -n_\text{max} \frac{k}{K}$,  and \eq{eq:ln:Z:pi} turns into
\begin{align}
\ln\big(Z_{\pi }\big) &= \ln q + n_\text{max}\frac{k}{K}\;.
\end{align} 
Based on our finding,  $n_\text{max} = \alpha \frac{N_{p} K}{k}$, we obtain
\begin{align}
\frac{\ln\big(Z_{\pi }\big)}{N_{p}} &= \frac{\ln q}{N_{p}} + \alpha\;,
\end{align} 
where $\alpha=O(1)$. We see that in the thermodynamic limit $\ln\big(Z_{\pi }(\beta )\big)\to \alpha N_{p}$, which is consistent with the requirement that $\ln\big(Z_{\pi }(\beta )\big)$ is an extensive quantity, as it is proportional to the free energy.
Moreover, it will be shown in this section, the relative statistical uncertainty of
$\ln(X_{n_\text{max}})$ is proportional to $1/\sqrt{N K}$, and therefore $Z_{\pi}$ can be determined for large systems from a single NESA run with high accuracy.

  \begin{figure}[h!]
	\centering
		\includegraphics[width=0.49\textwidth]{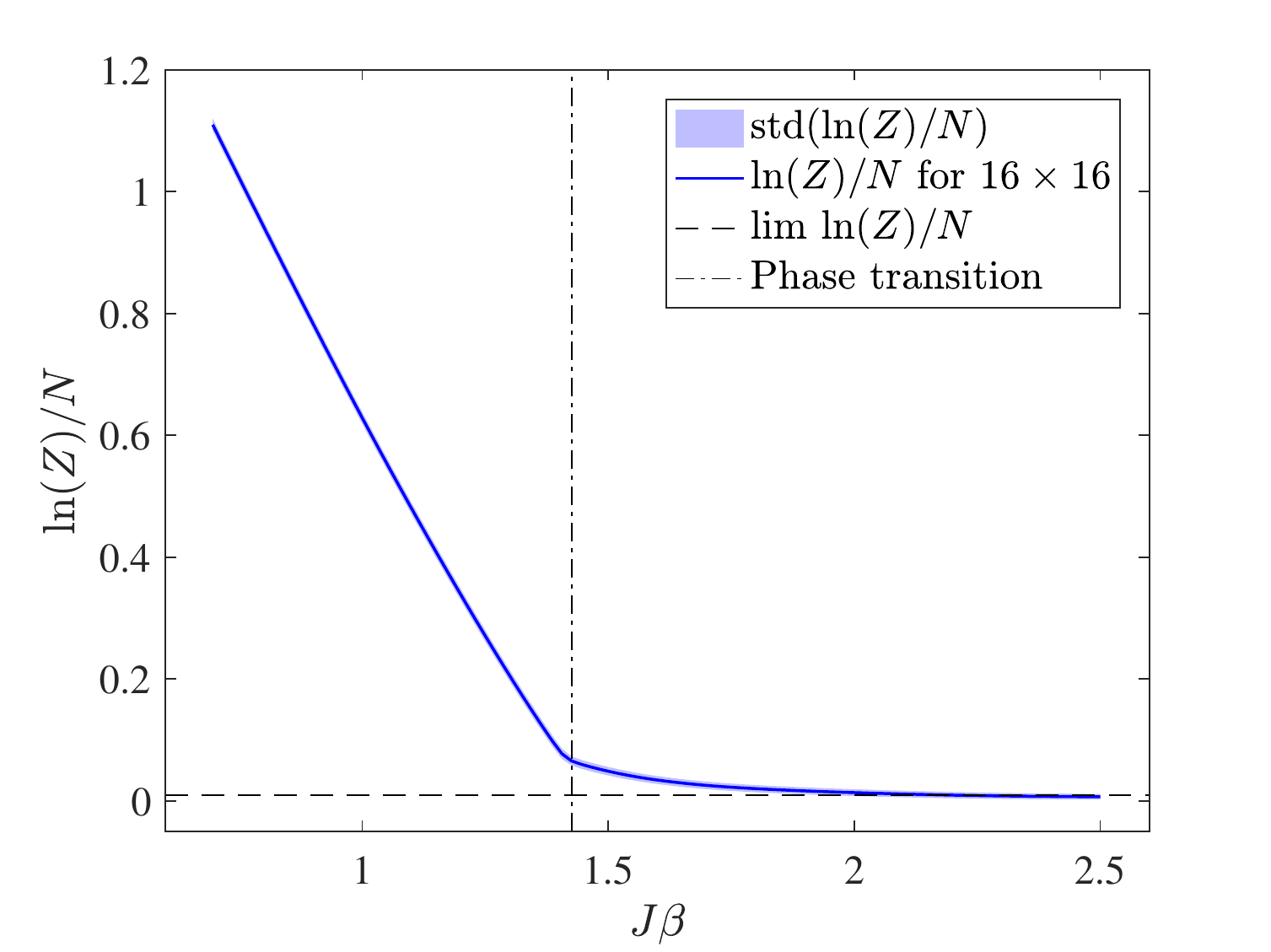}
		\includegraphics[width=0.49\textwidth]{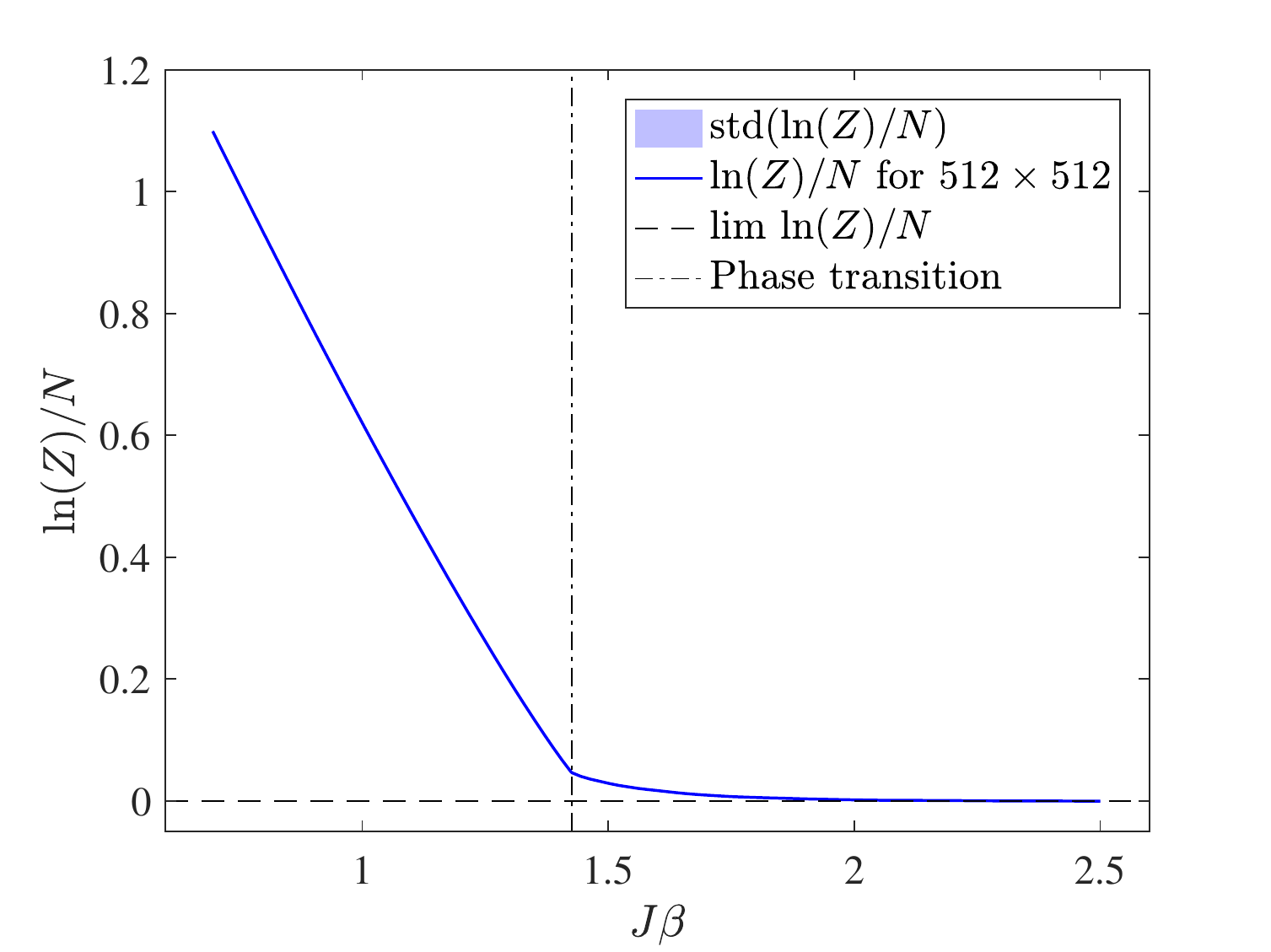}
	\caption{
	$\ln(Z)/N$  for a q=10 system versus $\betaJ$ evaluated via nested sampling with $K=100$ and $N_{\text{pr}} = 500$ (left panel) and $K=2$ and $N_{\text{pr}} = 500$ (right panel) walkers. The inverse critical temperature $\betaJ_{c}$ is marked by a vertical chain line and  the exact limit value for $\betaJ\to \infty$ (see \eq{eq:limits}) is represented by a dashed line. System sizes are $16\times 16$ (left panel)
	and 	$512\times 512$ 
	(right panel).
	}
	\label{fig:lnZ_16Jb}
\end{figure}

\Fig{fig:lnZ_16Jb} displays $\ln(Z)/N$ versus $\betaJ$ for the 10-state Potts model
for the system sizes $16\times 16$ and $512\times 512$.
For large systems, the statistical uncertainties are smaller than the line thickness.
The inverse temperature of the phase transition 
for an infinite 2d square lattice as given in \eq{eq:Tc:potts} is indicated
 by a vertical line.
 Above $\betaJ_{c}$, the curve approaches rapidly the limiting value of
 $\ln(Z)/N=\ln(q)/N$ for $\betaJ= \infty$, 
 which  is depicted by a dashed horizontal line. 
We observe that the largest uncertainty occurs at the lowest value of $\betaJ$, i.e. for $\betaJ^{*}$ which is
reasonable as the value at $\betaJ\to \infty$ was fixed to the exact result. 
$\betaJ^{*}$ corresponds to $\kappa=0$ and we have
\begin{align}
Z_{\text{P}}(\betaJ^{*}) &= Z_{\pi} e^{-\betaJ^{*} N_{p}} \underbrace{
\lim_{L\to \infty}\sum_{n=0}^{L}\Delta X_{n}
}_{\color{blue} = 1}\\
\ln\big(Z_{\text{P}}(\betaJ^{*})\big)&=\ln\big( Z_{\pi} \big) - \betaJ^{*}N_{p}\;.
\label{eq:kapa:0}
\end{align}
Hence, the accuracy of $\ln(Z_{\text{P}})$ at the inverse temperature $\betaJ^{*}$ is the same as that of $\ln(Z_{\pi})$, which is dictated by the 
distribution of $\ln(X_{n_\text{max}})$. 
Mean and variance of $\ln(X_{n_\text{max}})$ are derived 
 in appendix \ref{app:X:max}. Along with \eq{eq:n:max} for $n_\text{max}$ and \eq{eq:fixedseq10}  they are given by
\begin{subequations}\label{app:props:X:max}
\begin{align}
\bigg\langle \ln\big( X_\text{max} \big)\bigg\rangle 
&= \langle n_\text{max} \rangle\; \langle l_{1}\rangle\;
= L^{*}
\intertext{and}
\bigg\langle \bigg(\Delta \ln\big( X_\text{max} \big)\bigg)^{2}\bigg\rangle
&=2\;\langle n_\text{max}\rangle\;\langle \big(\Delta l_{1}\big)^{2}\rangle\;
=2\;L^{*} \;\frac{\langle \big(\Delta l_{1}\big)^{2}\rangle}{\langle l_{1} \rangle}
=\frac{2\;L^{*} \tilde k'}{\tilde k K} \; .
\end{align}
\end{subequations}
The  relative statistical uncertainty of $\ln(X_{n_\text{max}})$  is therefore
\begin{align}\label{eq:rel:uncertainty}
\varepsilon = \frac{\sqrt{\bigg\langle \bigg(\Delta \ln\big( X_\text{max} \big)\bigg)^{2}\bigg\rangle}}{\bigg|\bigg\langle \ln\big( X_\text{max} \big)\bigg\rangle \bigg|}&= 
\sqrt{\frac{2 \tilde k'}{L^{*} \tilde k K}}\;.
\end{align}
We recall that $L^{*} = -\ln( X_{n_\text{max}} )+O(k/K)$, as defined in \eq{eq:def:L:star},
and, therefore, we have
\begin{align}
\ln\big(Z_{\text{P}}({\betaJ^{*}})\big) &= \ln(q) -L^{*} + O(k/K)\;.
\end{align}
Since $\ln\big(Z_{\text{P}}({\betaJ^{*}})\big)$ is an extensive quantity, for large systems it has to be proportional to $N$ and, hence, \hbox{$L^{*}=f(q) N$}, with $f(q)$ being  a constant, independent of
$N$, $K$, and $k$, that will however depend on $q$. For the statistical uncertainty it 
means
\begin{align}\label{eq:rel:uncertainty:b}
\varepsilon = \frac{\sqrt{\bigg\langle \bigg(\Delta \ln\big( X_\text{max} \big)\bigg)^{2}\bigg\rangle}}{\bigg|\bigg\langle \ln\big( X_\text{max} \big)\bigg\rangle \bigg|}&= 
\sqrt{\frac{2 \tilde k'}{ f(q) \tilde k K N}}\;.
\end{align}
The pleasant bottom line is that the relative uncertainty is proportional to $1/\sqrt{N K}$, and it decreases with increasing system size.

\subsection{Internal energy}
Given $\ln(Z)$ as function of $\betaJ$ for a system in the canonical ensemble, thermodynamic quantities like the Helmholtz free energy F and the internal energy U 
as well as the entropy $S$ and heat capacity $c_V$  can be deduced \cite{pathria_statistical_2011}. 
Starting from the estimate of the partition function in \Eq{eq:ski_est}, 
expressions for the first and second derivative of $\ln(Z)$ w. r. t. $\betaJ$ can be determined. 
Given these expressions we can evaluate the physical quantities analytically, based on the sequence  of active bonds $\myhat{D}_{n}$ from a single NESA run and the sample of prior masses. Hence we can avoid the determination of numerical derivatives and the associated errors.
From \eq{eq:Liklins} we have 
 \begin{eqnarray} 
   \ln (L(\mathbf{b})) &=&  \kappa \;D(\mathbf{b})\;.
\end{eqnarray}
Given the sequence $\myhat{D}_{n}$ from a single NESA run and the corresponding prior masses $\Delta X_n$ we can calculate the logarithm of the partition function
\begin{eqnarray} \label{eq:summand}
  \ln (Z_\text{NESA}) &=& \ln  \big( \sum_{n} e^{ \kappa\myhat{D}_{n}}\: \Delta X_n\big) 
\end{eqnarray}
and the derivative with respect to $\betaJ$ 
\begin{align}
\begin{aligned}
\frac{\partial}{\partial\betaJ} \: \ln (Z_\text{NESA}(\betaJ)) 
&= \frac{1}{1-e^{-\betaJ}}\:
\sum_{n} \myhat{D}_{n}\:\underbrace{
\frac{ e^{ \kappa\myhat{D}_{n}}\: \:\Delta X_n}{Z_\text{NESA}}
}_{\color{blue} = p_{n}}\\
&= \frac{J}{1-e^{-\betaJ}}\:\avg{ D}_{\betaJ }\;.
   \end{aligned}
\end{align}
According to \eq{eq:Zns:a} we find e.g. for the internal energy
\begin{align}
U &= \avg{H} = -\frac{\partial}{\partial\betaJ} \: \ln (Z_{\text{P}}) 
= J \bigg(\Npair -  \frac{ \avg{D}_{\betaJ } }{1-e^{-\betaJ}}\bigg)\;.
\end{align}
The mean number of active bonds is related to the mean number of nearest neighbour pairs with equal spin  $\avg{N_\text{eq}(\mathbf{s})}_{\betaJ }$ via
$\avg{D}_{\betaJ } = p_\text{b} \avg{N_\text{eq}}_{\betaJ }$, with $p_\text{b}$ 
(see \eq{eq:p:active:bond}) being the probability that nearest neighbour pairs of equal spin form an active bond. 
Therefore, the internal energy can also be expressed as 
\begin{align}
	U &= - J\big( \avg{N_\text{eq}}_{\betaJ }-\Npair\big)\;,
\end{align} 
which is in agreement with the relation $U=-J\frac{\partial}{\partial \betaJ } \avg{H}$ and the definition of the Hamiltonian in \eq{eq:potts_energy}.
The second derivative can be deduced similarly.
Mean and variance of an observable $\mathcal{O}$, in the present case $U$, can be calculated using a single NESA run in a way similar to \eq{eq:moments:Z}. A single run produces $\vvmyhat{\lambda}$ which determines the observable $\mathcal{O}(\vvmyhat{\lambda},\vv \theta^{(m)})$ given the prior masses $\vv \theta^{(m)}$ drawn from \Eq{eq:beta:k}. Averaging over prior masses leads to
\begin{align} \label{eq:moments:O}
\avg{\mathcal{O}^{\gamma}} &= \frac{1}{N_{\text{pr}}}
\sum_{m=1}^{N_{\text{pr}}} \mathcal{O}(\vvmyhat{\lambda},\vv \theta^{(m)})^{\gamma}\;.
\end{align}

\begin{figure}[h!]
\centering
     \subfloat{
	      \includegraphics[width=0.6\textwidth]{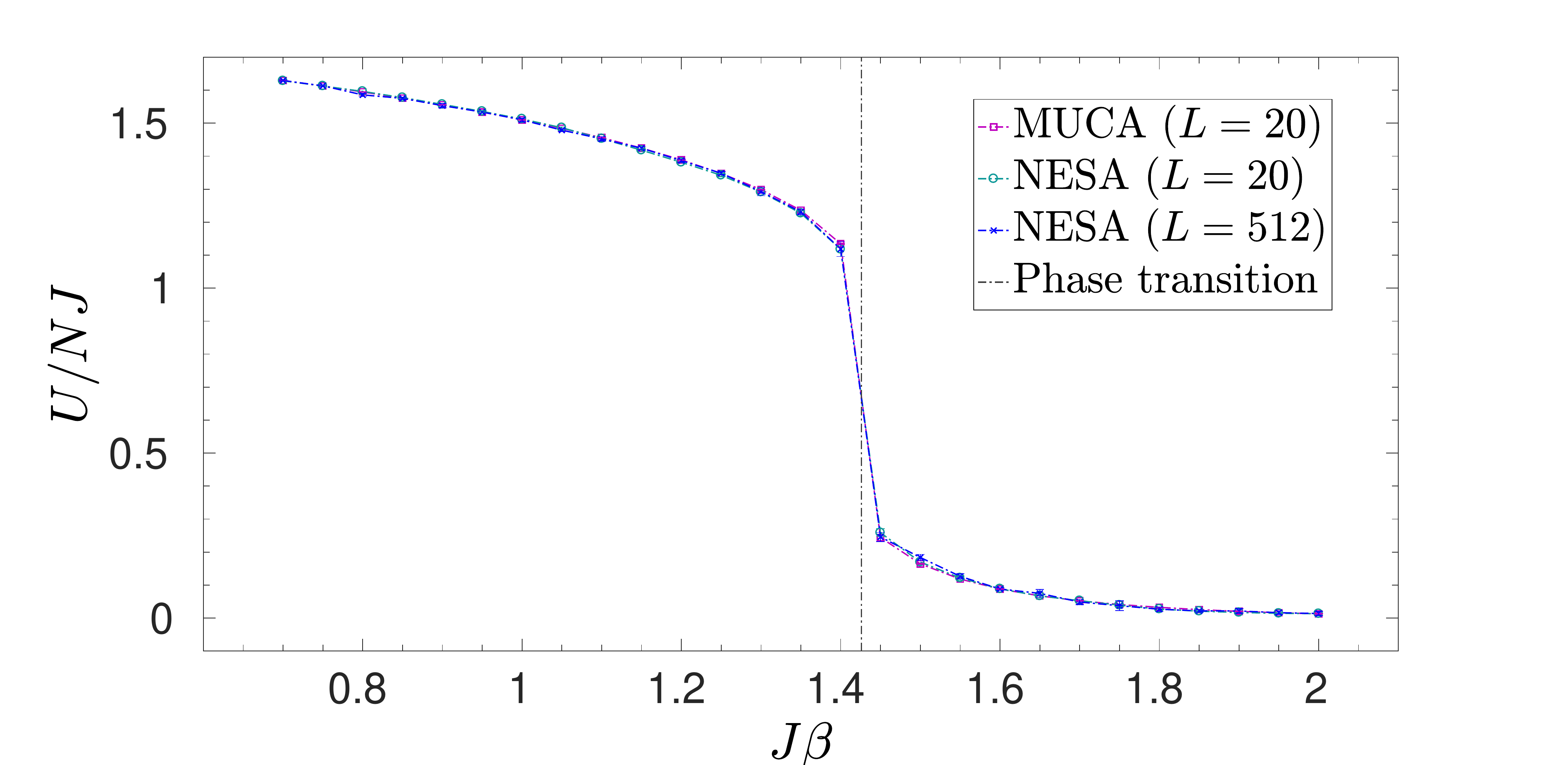}}
      \caption{Internal energy $U$ per site versus $\betaJ$ evaluated via nested sampling ($K = 500$, $N_{\text{pr}} = 500$) and via multi-canonical sampling for the $20\times20$ and $q=10$ and via nested sampling ($K=2$, $N_{\text{pr}} = 500$) for the $512\times512$ Potts system. The inverse critical temperature $\betaJ_{c}$ of the infinite system is marked by a vertical chain line.
      }	
      \label{fig:energy_muca_ns}
\end{figure}

\Fig{fig:energy_muca_ns} shows the internal energy $U$ versus $\betaJ$ for $q=10$ Potts systems computed via NESA and a MUCA simulation. For NESA at $L = 20$ $K = 500$ walkers and $N_{\text{pr}} = 500$ prior masses have been used and $K = 2$ and $N_{\text{pr}} = 500$ for the $L = 512$ system. The values agree excellently. For the MUCA simulation the Fortran code provided by Berg \cite{berg_multicanonical_2003} is employed. For the comparison of the NESA results to the MUCA results from \cite{berg_multicanonical_2003} the different definitions of $\betaJ$ and the Hamiltonian have to be taken into account.

\begin{figure}[h!]
	\centering
	  \subfloat{
		\includegraphics[width=0.6\textwidth]{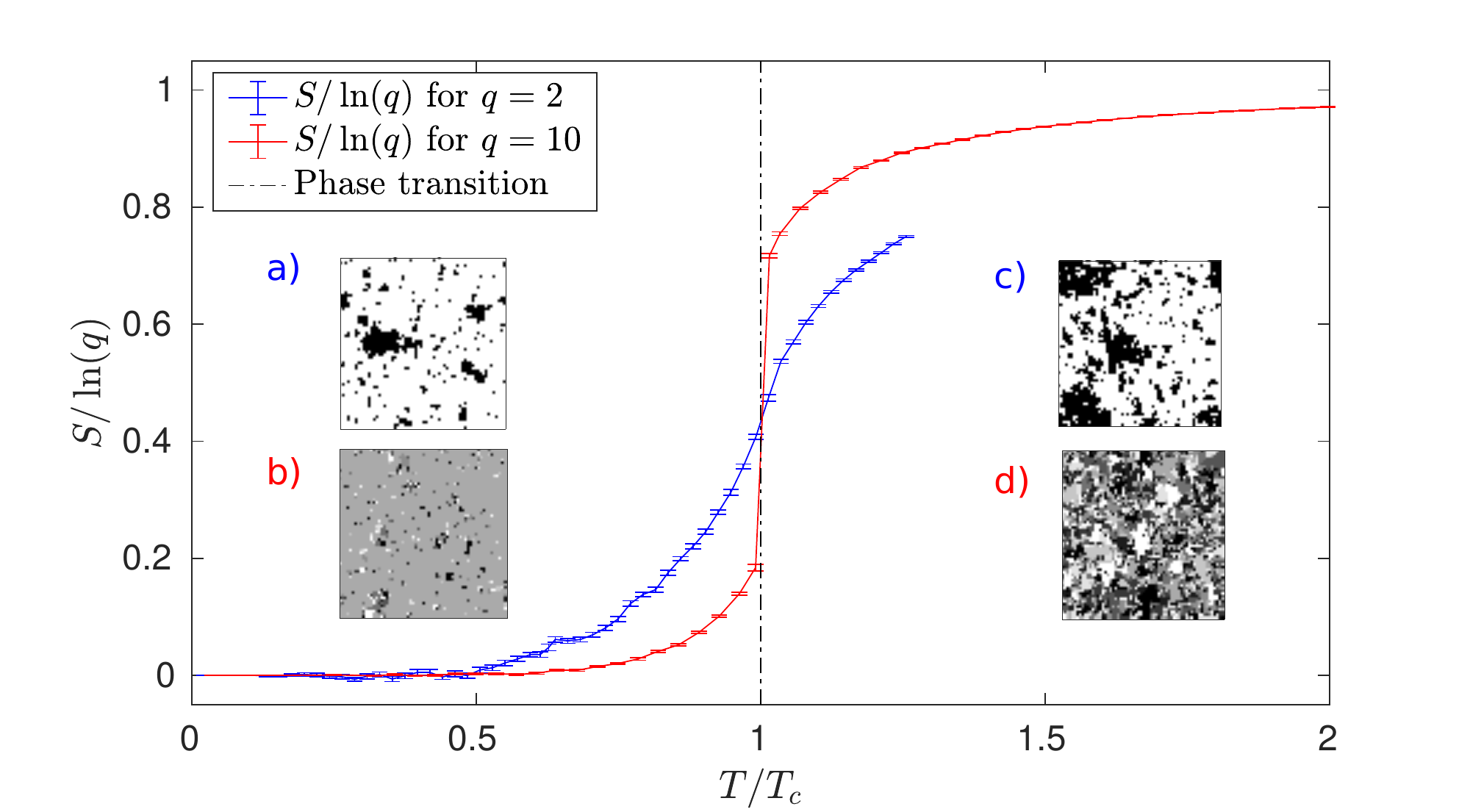}}
	\caption{Entropy for the $q=2$ (Ising-model) and $q=10$ Potts model at the $L=64$ system size calculated with $K = 512$ walker and $N_{\text{pr}} = 500$ prior masses. The snapshots of typical spin configurations are at $0.99 \; T_c$ for $q=2$ (a) and $q=10$ (b) and at $1.01 \;T_c$ for $q=2$ (c) and $q=10$ (d).}	
	\label{fig:entropy}
\end{figure}

The $q\leq 4$ Potts model exhibits a second order and the $q>4$ Potts model a first order phase transition. The difference in the order of the phase transition can be seen clearly in the discontinuity in the entropy
\begin{align}
    S = \beta U + \ln Z \;
\end{align}
in \Fig{fig:entropy}.
The snapshots in \Fig{fig:entropy} show typical spin configurations slightly below and above the critical temperature for the second ($q=2$) and the first ($q=10$) order phase transition. Whereas there is little structural change in the spin configuration near to the phase transition in the second order case, in the first order phase transition the configuration slightly above $T_c$ is fluctuating while below $T_c$ one cluster dominates the system.

\subsection{Magnetic properties and phase transition}

\begin{figure}[h!]
	\centering
	  \subfloat{
		\includegraphics[width=0.8\textwidth]{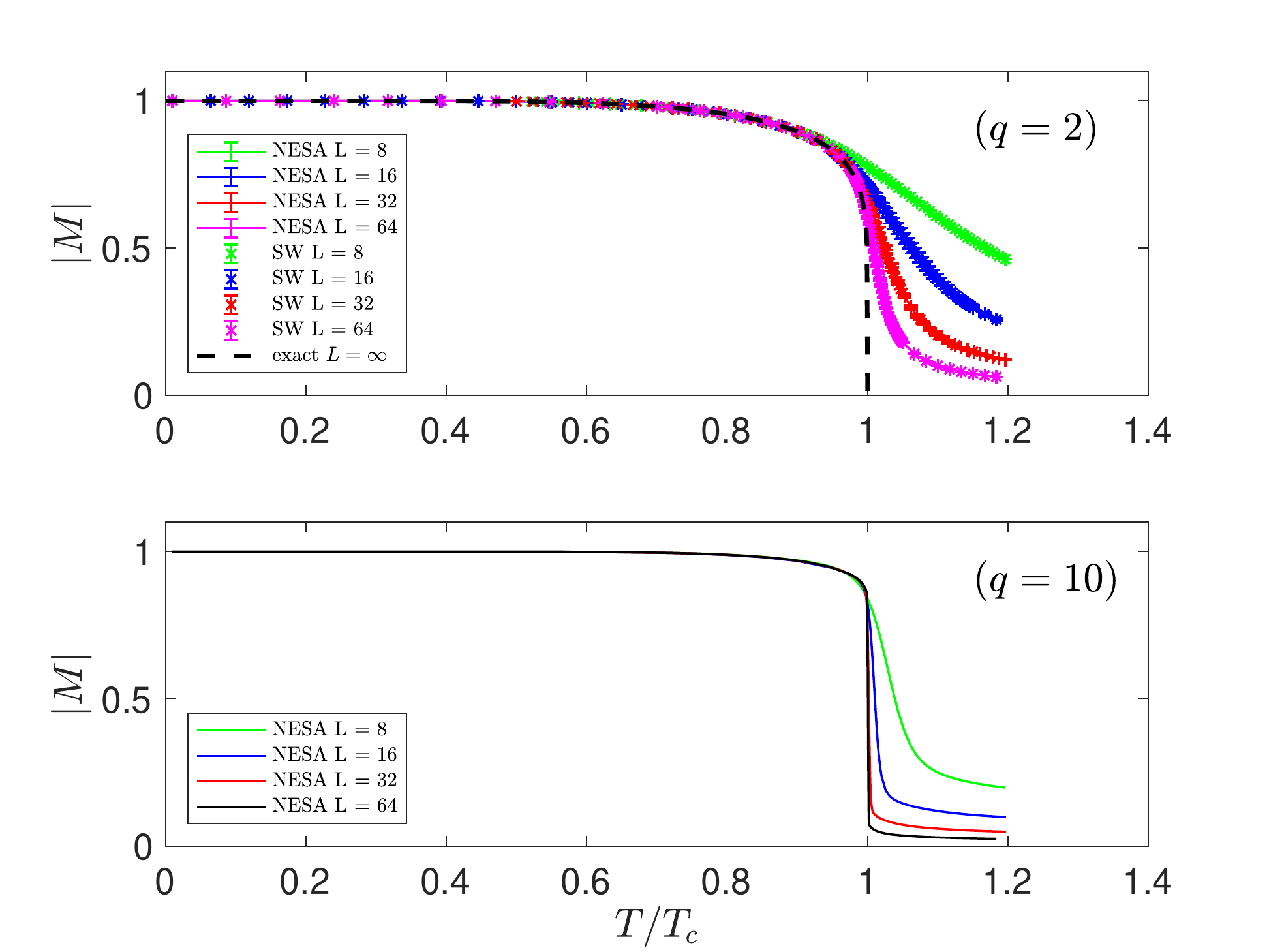}
		}
	\caption{
	Magnetization for the $q=2$ (upper panel) and $q=10$ (lower panel)  Potts model for different system sizes. For comparison, the exact result for the $q=2$ case for the infinite system is also depicted.
	System parameters are $K=1000, N_S=100, N_b=1, N_{\text{pr}}=1,N_{\text{avg}} = 10$.
	}
	\label{fig:magnetization}
\end{figure}

In order to compute magnetic properties of the Potts model we  add a magnetic field term 
\begin{align}\label{eq:H:B}
H_{B} &= -B\;{\cal M}(\vv s)
\end{align}
to the Hamiltonian. In this context, the magnetization of the spin configuration $\vv s$ is defined as
\begin{align}\label{eq:magnetization}
{\cal M}(\vv s) &= \sum_{i} e^{i \frac{2\pi}{q} s_{i}}\;.
\end{align}
The definition \eq{eq:magnetization} has  familiar  limiting values.  
In the high temperature limit, where all spin-configurations have the same probability, we obtain
\begin{align}\label{eq:M:B:zero}
\langle \mathcal M \rangle_{\beta=0,B= 0} &= 0\;\qquad \text{for any system size } N,\\
\frac{1}{N}\langle \big|\mathcal M \big|\rangle_{\beta=0,B= 0} &= 0\;\qquad \text{for } N\to \infty\;.
\end{align}
The last equation follows from \eq{app:M:limit} in appendix \ref{app:hightempM}.
In the opposite limit, $T \to 0$, all spins are equal. Here we have to distinguish the two observables. For
${\cal M}$, we need to perform the thermodynamic limit $N\to \infty$ before we reduce $B\to 0$. Then we find 
\begin{align}\label{eq:M:T:zero:M}
\langle {\cal M} \rangle_{N\to \infty, T=0,B\to 0} &= N\;.
\end{align}
In case of the modulus, we can take $B=0$ also for a finite $N$ and obtain
\begin{align}\label{eq:M:T:zero}
\langle \big| {\cal M} \big| \rangle_{N,T=0,B= 0} &= N\;.
\end{align}
The magnetization $\langle \big|{\cal M}\big| \rangle/N$ for $q=2$ and $q=10$ is depicted in \Fig{fig:magnetization} for various system sizes.
For the Ising model ($q=2$) we compare the NESA results to results obtained with a Markov Chain Monte Carlo algorithm with Swendsen Wang (SW) updates. For the infinite Ising model Onsager's formula gives the exact solution for the susceptibility below $T_c$ which fits well to our calculations.
The high temperature limits for the magnetization are derived in appendix \ref{app:hightempM}.
The behavior of the magnetization at the phase transition shows that the $q=2$ Potts model exhibits a second order and the $q=10$ Potts model a first order phase transition.

\begin{figure}[h!]
	\centering
	  \subfloat{
		\includegraphics[width=0.99\textwidth]{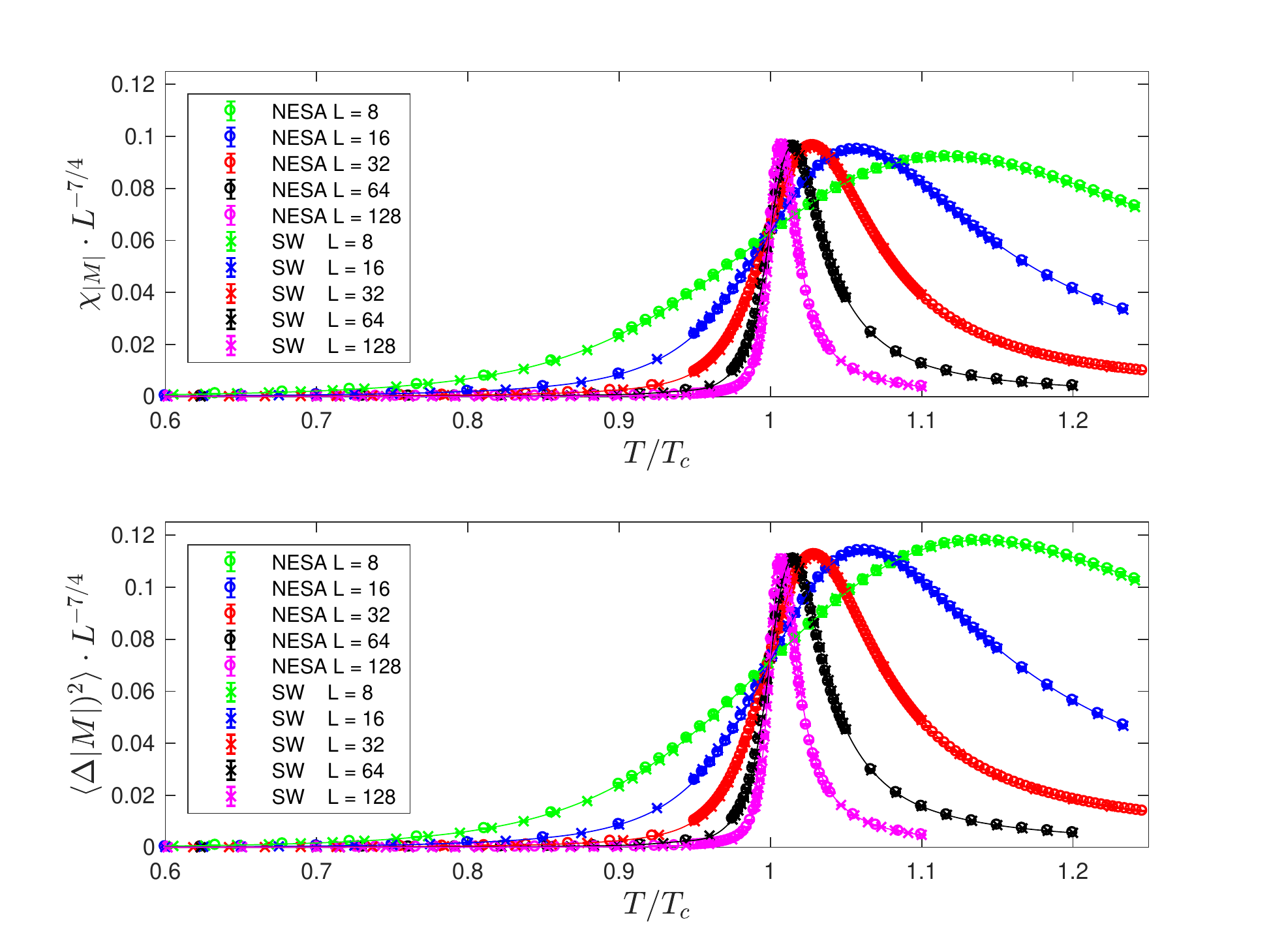}}
	\caption{Comparison of the susceptibility for the $L = \{8,16,32,64,128\}$ Ising model as obtained by NESA and Swendsen Wang. The scaling factor $L^{\gamma/\nu}$ is already included and we have also plotted the variance of the magnetization . We recognize the correct scaling in the fact that the height of the peaks does not change for larger systems.
	System parameters are $K=1000, N_S=100, N_b=1, N_{\text{pr}}=1,N_{\text{avg}} = 10$.}	
	\label{fig:susceptibility}
\end{figure}

The thermodynamic expectation value of functions $f({\cal M})$ of the magnetization are defined by
\begin{align}\label{eq:M:avg}
\langle f({\cal M}) \rangle_{\beta,B} &= \frac{1}{Z} \sum_{\vv s} e^{\betaJ \sum_{\langle ij\rangle}
\delta_{\sigma_{i},\sigma_{j}} + \beta B {\cal M}(\vv s) } \;f({\cal M}(\vv s))\;.
\end{align}
The computation of expectation values in the framework of NESA in the Potts model is  outlined in appendix \ref{app:expectationValue:Potts}. 
The standard expression for the susceptibility is
\begin{align}                                    \label {MCMC:SW:susz4}
  \chi(\beta,B)&= \frac{\beta}{N}\bigg(
\avg{{\cal M}^2} -
\avg{ {\cal M}}^2 \bigg)\;.
\end{align}
It   has the drawback that for finite systems  the magnetization vanishes for $B\to 0$
and, therefore, the susceptibility does not decrease for $T$ below $T_{c}$.
A more suitable estimate for the susceptibility for finite systems is given by \cite{janke_monte_1996}
\begin{align}                                    \label{MCMC:SW:susz:modulus}
  \chi_{|{\cal M}|}(\beta,B) &= \frac{\beta}{N}\bigg(
\avg{{\cal M}^2 } -
\avg{ |{\cal M}|}^2 \bigg)\;.
\end{align}
Finally we plotted the susceptibility $\chi_{|{\cal M}|}$ for the $q=2$ Potts model for the system sizes $L\in\{8,16,32,64,128\}$ in \Fig{fig:susceptibility} and compare the nested sampling results with those obtained by the Swendsen Wang
algorithm. First of all we observe a perfect agreement.
Moreover, as expected, with increasing system size the resonance becomes narrower and is getting closer to the critical temperature of the infinite system, both scale as $1/L$. 
We have scaled the susceptibility with $L^{\gamma/\nu}$ to show that the
maximum values $\chi^{max}_{|{\cal M}|}/L^{\gamma/\nu}$ stay constant for larger system sizes for a critical exponent of $\gamma/\nu=7/4$, in agreement with the known finite size scaling, which is
 discussed in detail in \cite{newman_monte_1999,janke_monte_1996}.
 Next we turn to the critical exponent $\gamma/\nu$ that governs the behaviour of $\chi_\text{max}$, the maximum of susceptibility. We can either perform a non-linear fit of the maxima of $\chi$ obtained by NESA
 for the system sizes $L\in\{8,16,32,64,128\}$ with $\chi^{max}_{|{\cal M}|}\propto L^{\gamma/\nu}$.
 This yields $\gamma/\nu = 1.754\pm 0.002$, while the fit of  the maxima of  $\avg{(\Delta |M|)^{2}}$ yields $\gamma/\nu = 1.735\pm 0.002$. It should be pointed out that the scaling law 
 $\propto L^{\gamma/\nu}$ can still contain finite size corrections of the form $\propto L^{\gamma/\nu}(1+O(1/L))$. These corrections are clearly visible in figure \ref{fig:susceptibility} and it can also be seen with the naked eye that the finite size correction is more pronounced in $\avg{(\Delta |M|)^{2}}$.
 To eliminate the finite size corrections, we determining the slope of $\ln(\chi_\text{max}(L)) = a + b \ln(L)$ 
 for neighbouring system-sizes, plot the slope as function of $1/L$ and extrapolate to $1/L\to 0$. This results in $1.75\pm 0.01$ and $1.74\pm 0.02$, for $\chi$ and $\avg{(\Delta |M|)^{2}}$
 respectively.

In addition to the susceptibility $\chi = \beta \avg{(\Delta |M|)^{2}}$, we have also depicted the factor $\avg{(\Delta |M|)^{2}}$. Both terms have the same critical exponent $\gamma/\nu$ and differ only by the the way they approach the thermodynamic limit.
It is clearly visible from the figure that  $\chi$ approaches the thermodynamic limit of $\gamma/\nu$ from above, i.e. the height of the peaks in the upper plot monotonically increase with increasing system size, but the increase drops to zero with $N\to \infty$.
This is in accord with the numerical value given above. On the other hand,
$\avg{(\Delta |M|)^{2}}$ approaches the thermodynamic limit from below.
As defined in the appendix \ref{app:expectationValue:Potts}, there are two sample sizes $N_{b}$ and $N_{S}$, which can both be set to 1, but in this case the statistical uncertainty has to be determined from repeated
nested sampling runs. 
In this application it turned out to be favourable to replace the prior masses $X_{m}$ by the mean values
and to estimated the statistical noise by repeated NESA runs, which was necessary in any case to account for 
the statistical uncertainty introduced by the averages discussed in appendix \ref{app:expectationValue:Potts}.
Otherwise, in all runs for figure \fig{fig:susceptibility} we have used $K=1000$ and $N_{b}=1$ and $N_{S}=100$. 
The later is necessary to reduce the statistical noise and it is faster than 100 repeated nested sampling runs.
Then it occurs that the statistical noise is very small and the confidence intervals are determined from $N_{\text{avg}}$ 
repeated NESA runs. The reason why $N_{b}$ is already sufficient is due the the fact that in case of $K=1000$, the plateaus in $D_{n}$ in \fig{fig:D_n} are roughly $K$, which is equivalent to using $N_{b}=1000$.

\subsection{Performance comparison multi-canonical sampling and nested sampling} \label{sec:compnsti}

\begin{figure}[h!]
	\centering
	  \subfloat{
		\includegraphics[width=0.6\textwidth]{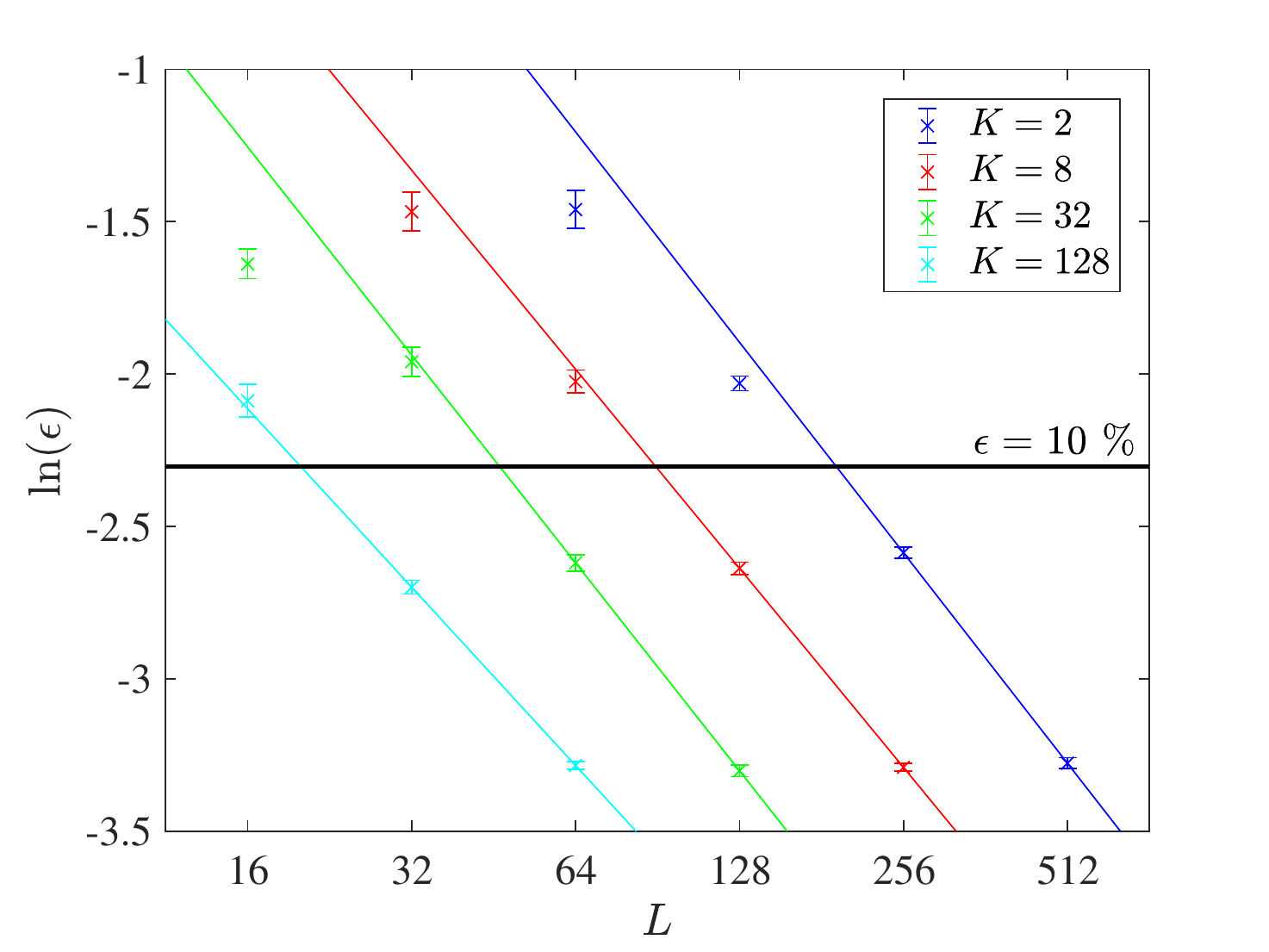}}
	\caption{Logarithm of the relative accuracy of $\ln(Z)/N$ for the q=10 Potts model at $\betaJ=\betaJ_{c}$ computed by nested sampling in dependence of the logarithmic grid-size and the number of walker $K$.
	}	
	\label{fig:TI_lnZ_size}
\end{figure}

For a proper comparison of MUCA and NESA we analyse the computing times needed for the calculation of $\ln(Z)/N$ with a defined relative accuracy for the $q=10$ Potts model at the critical temperature $\betaJ_{c}$.
As derived in \Se\ref{sec:stoch:eval:Z} the asymptotic behaviour of the relative accuracy in NESA is proportional to $1/\sqrt{N K}$. \Fig{fig:TI_lnZ_size} shows the computed logarithmic relative uncertainty depending on the system size and the number of walker. The black solid line marks the relative accuracy of $10~$\%.
For each number of walker we do a line fit only for the data points below the $10~$\%-line, because there the asymptotic behaviour should hold.
We get nearly equidistant lines with a distance of $\Delta \ln (\epsilon) \approx 0.69$ and a slope of approximately $-0.5$ which confirms the scaling law derived analytically.

\begin{figure}[h!]
\centering
	  \subfloat{
		\includegraphics[width=0.6\textwidth]{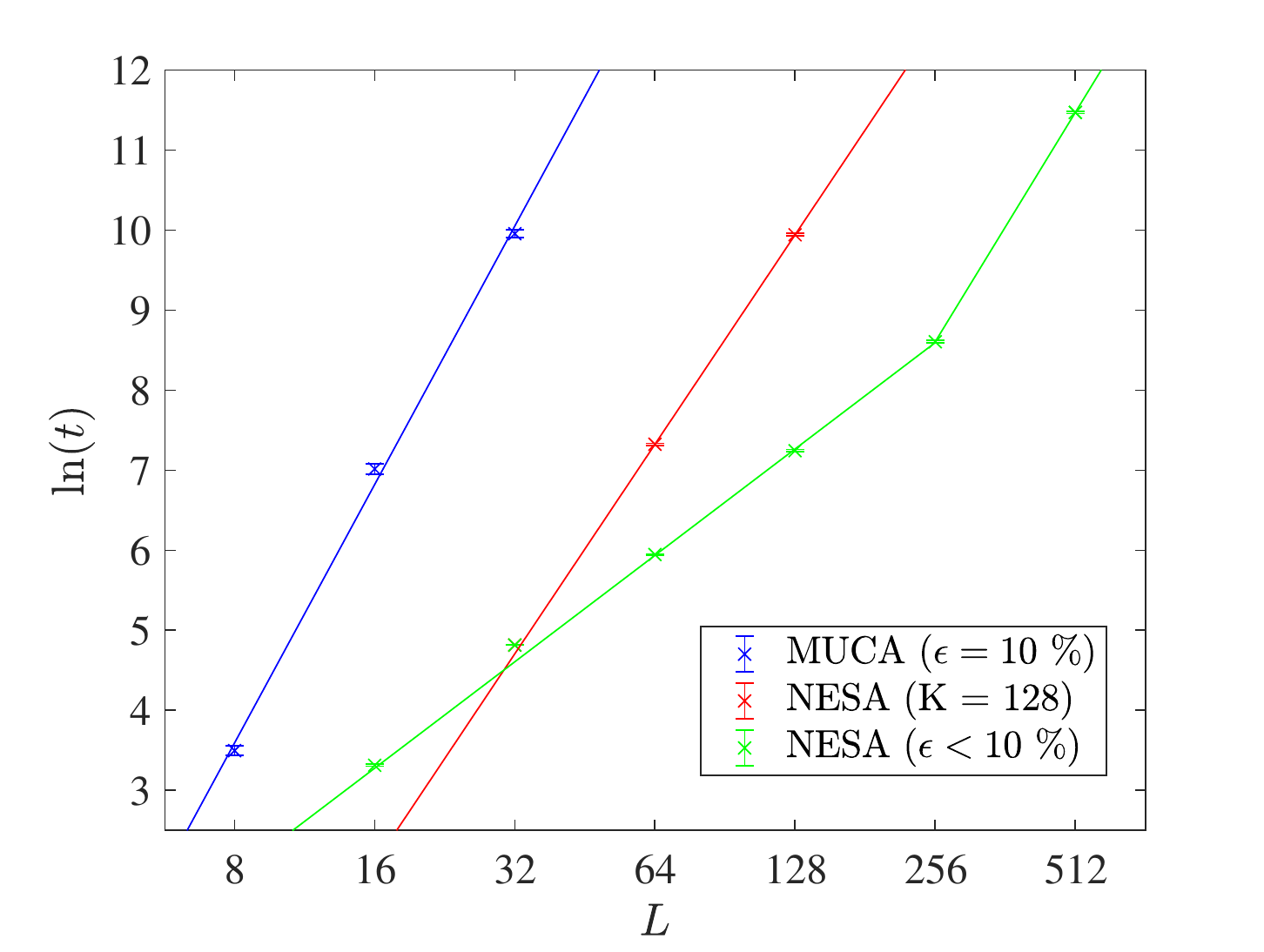}}
	\caption{Logarithmic time for computing $\ln(Z)$ versus the logarithmic grid size for the q=10 Potts model at $\betaJ=\betaJ_{c}$ exhibits a slightly stronger scaling for MUCA than for NESA.
	}	
	\label{fig:TI_lnZ_size222}
\end{figure}

For the comparison of MUCA and NESA we analyse the scaling exponent $x$ of the computing time $t \propto N^x$ needed for calculating $\ln(Z)/N$ with a relative accuracy of at least $10~$\%.
In MUCA we used a sample of 100 independent walker and stopped the computation when the requested relative accuracy is reached.
We have implemented the MUCA simulation using local updates and found it scales with $x = 2.36 \pm 0.11$, see blue line in \fig{fig:TI_lnZ_size222}, which is consistent with the scaling of $x \approx 2.3$ obtained in \cite{berg_introduction_1999} for the same system. Here we have neglected the time needed for computing the weights needed for MUCA which scales with $N^2$.
In NESA we found that the computing time scales with $x \approx 2$ if we use a constant number of walker for every system size, see red line in \fig{fig:TI_lnZ_size222}. The scaling can be explained by the SW updates needed in each step scaling with $kN$ and the maximum number of steps scaling with $KN/k$, as discussed in section \ref{ssec:z_and_ddz}. Hence, the total time scales like $t \propto KN^2$.
But, as pointed out before, decreasing the number of walker for increasing system sizes produces the same relative accuracy. Therefore we have a linear scaling in this case, $t \propto \epsilon^2 N$. We performed calculations for $L = \{ 16, 32, 64, 128, 256, 512 \}$ at $K = \{ 256, 128, 32, 8, 2, 2\}$ which ensures that the relative accuracy is lower than $10~$\% and we get a linear scaling in the region where $K$ can be decreased and a quadratic scaling otherwise, see green line in \fig{fig:TI_lnZ_size222}.
Additional to the smaller scaling exponent of NESA also the prefactor is smaller compared to MUCA.

\subsection{Parallel nested sampling}\label{ssec:parallel_ns}
As outlined in \Se\ref{sec:stoch:eval:Z} the greatest uncertainty occurs at 
$\betaJ^{*}=\ln(2)$ and the corresponding relative uncertainty is given in
\eq{eq:rel:uncertainty:b}.
For $k\ll K$ we have $\tilde k = \tilde k'$ and then
\begin{align}\label{eq:}
\varepsilon&= \sqrt{\frac{2 }{f(q) NK}}\;.
\end{align}
Hence, the relative uncertainty behaves like $1/\sqrt{N K}$, independent of $k$, which is corroborated by our numerical results shown in \fig{fig:TI_lnZ_size}.
This result deviates from the conjecture of Henderson \etal{} \cite{henderson_parallelized_2014}
that the uncertainty increases with $\sqrt{k}$, based on the  increasing variance 
of a single shrinkage factor $\langle (\Delta l_{1})^{2} \rangle$. This is fortunately not the case for the present application. As the uncertainty is independent of $k$,  NESA
is perfectly suited for a parallel implementation. The $k$ walkers are drawn independently from each other according to the prior and this step consumes for large systems the lion's share for the CPU time.

\section{Summary and conclusions}\label{sec:conclusion}

In this work we have evaluated the nested sampling (NESA) algorithm in the frame of the Potts model on a 2d square lattice  for different system sizes and numbers of possible spin values $q$. Results and performance are  compared  with the established multi-canonical (MUCA) sampling method. We have employed the Fortuin and Kasteleyn bond representation. The primary goal was the computation of the partition function, which is difficult to determine reliably by standard Monte Carlo techniques.
Both methods, multi-canonical sampling and nested sampling, exhibit a power law scaling of the computation time with increasing system-size. We find an exponent of $2$  for NESA, while it is $\approx 2.3$ for MUCA, demonstrating the  superiority of nested sampling. Moreover, in NESA the relative uncertainty of the logarithmic partition function scales with $1/\sqrt{N}$, which allows to obtain very accurate results for large system sizes.
Another advantage of nested sampling is that it can easily be implemented and does not require additional adjustments, like the  weights $w(E)$ in MUCA, which becomes increasingly cumbersome with increasing system size.

Besides the evaluation of the partition function, we have demonstrated that it is possible to directly compute thermodynamic expectation values with NESA, both for observables that can be expressed entirely in the bond representation (improved estimators) and those that still need the original spin representation. Results were given for the internal energy, entropy, Helmholtz free energy, magnetization and
magnetic susceptibility, for which we determined the critical exponent for the Ising model. In all cases we
we found excellent agreement between the NESA results and those obtained by MUCA or the Swendsen-Wang algorithm.

In addition nested sampling is perfectly suited for parallel computing.
In summary we have found that nested sampling is able to deal efficiently with 
problems that exhibit first order phase transitions. 
The method is a promising alternative to the multi-canonical algorithm, which is a state of the art computational technique for dealing with Potts-type of models. 
In our opinion nested sampling represents a high potential for applications in statistical physics and due to its uniqueness it deserves a place in a physicists standard repertoire of simulation techniques.

\section{Appendix}
\subsection{Nested sampling the partion function\label{app:NESA:Z}}

For the computation of integrals or sums of the form
\begin{align}\label{eq:}
Z &= \SumInt{x} L(\vv x) \pi(\vv x)\;,
\end{align}
where $\pi(\vv x)$ represents  a probability function,
we introduce the dummy integral $\int d \lambda\delta(L(\vv x) - \lambda)=1$ and obtain 
after swapping the integrals
\begin{align}\label{eq:Z:app1}
Z &=  \int d\lambda \;\lambda \; \SumInt{x} \; \pi(\vv x)\;\delta(L(\vv x)-\lambda)\;.
\end{align}
The inner integral is related to the prior mass $X(\lambda)$, which is defined as
\begin{align}\label{app:prior:mass}
X(\lambda) &=  \SumInt{\vv x}	 \; \pi(\vv x)\;\theta(L(\vv x)-\lambda)\;,
\end{align}
via
\begin{align}
\SumInt{\vv x} \; \pi(\vv x)\;\delta(L(\vv x)-\lambda) &= -\frac{d}{d\lambda} X(\lambda)\;.
\end{align}
Inserting this relation into \eq{eq:Z:app1} yields
\begin{align}\label{eq:Z:app2}
Z &=  -\int  \lambda \; \frac{d X(\lambda)}{d\lambda}\;d\lambda
=  \int  {\cal L}(X) \;d X\;,
\end{align}
where ${\cal L}(X)$ is the inverse function of $X(\lambda)$. We have tacitly assumed that 
$X(\lambda)$ and ${\cal L}(X)$ are strictly monotonic functions. From the definition of the prior mass
$X(\lambda)$ it is clear that it decreases monotonically with increasing $\lambda$.
Therefore an additional minus sign appears in the last step due to switching the integration limits.
It is, however, not always
automatically ensured that $X(\lambda)$ is strictly monotonic, but it can easily be added.

\subsection{Nested sampling discrete configuration spaces}
\label{sec:NESA:details}
\begin{figure}[b!]
\centering
\includegraphics[width=0.75\textwidth]{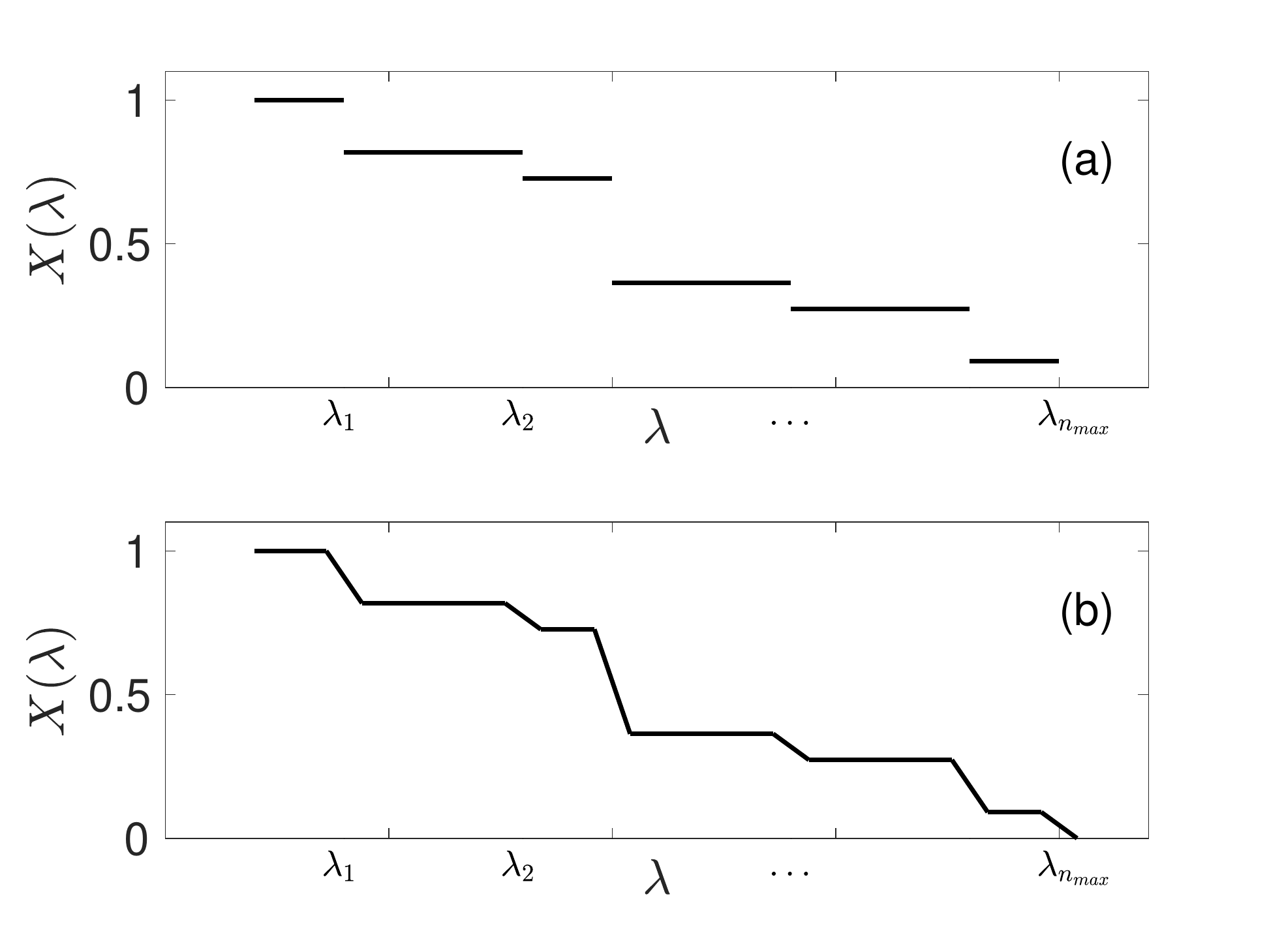}
\caption{Prior masses $X(\lambda)$ versus likelihood value $\lambda$ for the purely discrete case configuration space (a) and the augmented case (b).\label{fig:stairway}} 
\end{figure}

Here we will analyse some details of the algorithm outlined in section \ref{ssec:nesa}.
In case of a discrete problem, some additional considerations are in order. We consider a discrete
configuration space $\vv x$. The corresponding likelihood $L(\vv x)$ values are also discrete and 
it can have the following values $L_{\nu}$. The corresponding prior masses are
\begin{align*}
X(\lambda) 
&=\sum_{\vv x}\; \pi(\vv x)\;\theta(L(\vv x) > \lambda)\;\underbrace{
\sum_{\lambda_{\nu}} \delta_{L(\vv x),\lambda_{\nu} }
}_{\color{blue} = 1}\;.
\end{align*}
\begin{align}
X(\lambda) &=\sum_{\lambda_{\nu}}\;P(\lambda_{\nu}) \;\theta(\lambda_{\nu} > \lambda)\\
P(\lambda_{\nu})&=\sum_{\vv x}\; \pi(\vv x)\; \delta_{L(\vv x),\lambda_{\nu} }
\;.
\end{align}
Clearly, $X(\lambda)$ is  a discontinuous (open) stairway function as depicted in figure \ref{fig:stairway} and, therefore, the PDF of the prior mass is far from being uniform. 
However, we show that the augmented  configuration space introduced in section \ref{ssec:NESAalgorithm} to overcome 
degeneracy of the likelihood by an additional variable $m$ with uniform PDF
\begin{align}
p(m) &= \frac{1}{2 \varepsilon}\theta(-\varepsilon \le  m <  \varepsilon)\;
\end{align}
can be used to obtain a continuous $X(\lambda)$.
The augmented  likelihood is $L(\vv x,m) = L(\vv x) + m$.
Next we compute the PDF for the likelihood values that occur in the algorithm with the augmented
configurations
\begin{align*}
p(\lambda\CS\BK)  &= \sum_{\vv x }\int dm \;p(m) \pi(\vv x) \underbrace{
p(\lambda\CS\vv x, m,\BK)
}_{\color{blue} = \delta(\lambda - L(\vv x)-m)}\\
 &=\sum_{\lambda_{\nu}} \bigg(\sum_{\vv x }\pi(\vv x) \delta_{L(\vv x),\lambda_{\nu}}\bigg)\;
\bigg(\int dm p(m) \delta(\lambda - \lambda_{\nu}-m)\bigg)
\end{align*}
resulting in 
\begin{align}
p(\lambda\CS\BK)&= \sum_{\lambda_\nu} \frac{P(\lambda_{\nu})}{2\varepsilon} \;\theta(\lambda\in I_{\nu})
\end{align}
with the definition of the intervals
\begin{align}
I_{\nu} &= [\lambda_{\nu}-\varepsilon, \lambda_{\nu}+\varepsilon)\;.
\end{align}
Hence $p(\lambda|\BK)$ is a sum of bars centred at the positions  $\lambda_{\nu}$ with height $P(\lambda_{\nu})/2\varepsilon$ and width $2\varepsilon$.
Now we compute the prior mass 
\begin{align*}
X(\lambda) 
&=\sum_{\vv x}\; \pi(\vv x)\;\int  dm \;p(m)\;\theta(L(\vv x) - \lambda> m)\;
\underbrace{
\sum_{\lambda_{\nu}} \delta_{L(\vv x),\lambda_{\nu} }
}_{\color{blue} = 1}\;,
\end{align*}
which leads to
\begin{align}
X(\lambda) &=\sum_{\lambda_{\nu}}P(\lambda_{\nu})\;W_{\nu}(\lambda)\;.
\end{align}
The window function $W$ is given by
\begin{align}
W_{\nu}(\lambda)&=\int dm \; p(m) \;\theta(m< \lambda_{\nu}-\lambda)  \notag \\
&= \int dm \; \frac{1}{2\epsilon} \theta(-\epsilon < m< \epsilon) \;\theta(m< \lambda_{\nu}-\lambda) \notag \\
&= \theta(\lambda< \lambda_\nu-\epsilon) \frac{1}{2\epsilon} \int_{-\epsilon}^{\epsilon} dm \;  + \; \theta(\lambda \in I_{\nu}) \frac{1}{2\epsilon} \int_{-\epsilon}^{\lambda_\nu-\lambda} dm \; \notag \\
&= \theta(\lambda< \lambda_\nu-\epsilon) \; + \; \theta(\lambda \in I_{\nu})\;	\frac{\lambda_{\nu}-(\lambda-\varepsilon)}{2\varepsilon}
\end{align}
Now $X(\lambda)$ is a continuous function in $\lambda$ which looks like the original stairway but the 
risers are closed by straight lines with the finite slope $P(\lambda_{\nu})/2\varepsilon$, see  figure \ref{fig:stairway}.
Now $\frac{d X(\lambda)}{d\lambda}$
is finite everywhere, which is important for the following considerations. Moreover, we see that 
\begin{align}\label{app:dXdl}
\frac{d X(\lambda)}{d\lambda} &= -p(\lambda\CS\BK) \;;\quad\forall \lambda \in \cup_{\nu} I_{\nu}\;. 
\end{align}
Now we can compute the PDF of $X$, where we exploit that $p(\lambda\CS\BK)$ is non-zero only for
$\lambda \in \cup_{\nu} I_{\nu}$  
\begin{align*}
p(\tilde X\CS \BK) &= \int d\lambda \underbrace{
p(\tilde X\CS \lambda,\BK)
}_{\color{blue} = \delta(\tilde X -X(\lambda))} p(\lambda\CS\BK)\\
 &= \sum_{\nu} \int_{I_{\nu}} d\lambda 
 \delta(\tilde X -X(\lambda)) p(\lambda\CS\BK)\\
 &= \sum_{\nu} \int_{I_{\nu}} d\lambda 
\frac{ \delta(\lambda - {\cal L}(\tilde X))}{|\frac{dX(\lambda)}{d\lambda}|} p(\lambda\CS\BK)\\
 &= \sum_{\nu} \int_{I_{\nu}} d\lambda \delta(\lambda - {\cal L}(\tilde X))=\theta(0\le \tilde X\le 1)\;.
\end{align*}
In the last step we have used \eq{app:dXdl}.

\subsection{Distribution of $n_\text{max}$\label{app:n:max}}

Assuming we know ${\cal L}(X)$ and its inverse, then we can determine $X^{*}$ defined as
the prior mass with
\begin{subequations}\label{eq:def:L:star}
\begin{align}
{\cal L}(X) &= {L}_\text{max}\;\quad\forall X\le X^{*}\;,\\
{ L}^{*}&:=-\ln(X^{*}).
\end{align}
\end{subequations}
Based on \eq{eq:X:max} we have for large $N$ that ${ L}^{*}\propto N$.
To compute $P(m\CS \BK):=P(n_\text{max}=m\CS \BK)$ we 
introduce the set $\vv l = \{l_{\nu}\}$
of logarithmic shrinkage factors, from which we actually only need $\nu=1,\ldots, m+1$,
\begin{align}
P(m \CS \BK) &= 
\int \bigg(\prod_{\nu=1}^{m+1} d l_{\nu}\bigg)
P(m \CS  \vv l,\BK) \prod_{\nu}^{m+1} p(l_{\nu}).
\end{align}
We have exploited the fact that the shrinkage factors are independent random variables.
Now the probability $P(m \CS  \vv l,\BK)$ is simply
\begin{align}
P(m \CS  \vv l,\BK) &= 
\begin{cases}
	1&\text{if } \sum_{\nu=1}^{m} l_{\nu} < { L}^{*}  \;\wedge \sum_{\nu=1}^{m+1} l_{\nu} > { L}^{*} .\\
	0&\text{otherwise} 
\end{cases}
\end{align}
We introduce the variable $S_{m}:=\sum_{\nu=1}^{m}l_{\nu}$ by marginalization
\begin{align}
P(m \CS \BK) &= \int dS_{m}
\int \bigg(\prod_{\nu=1}^{m+1} d l_{\nu}\bigg)
\theta\bigg( 
S_{m}< { L}^{*}  \;\wedge S_{m}+ l_{m+1}  > { L}^{*} 
 \bigg)
\underbrace{
 p(S_{m}\CS \vv l)
}_{\color{blue} = \delta(S_{m}=\sum_{\nu=1}^{m} l_{\nu})}
\prod_{\nu}^{m+1} p(l_{\nu})\notag \\
&=\int dS_{m}\int dl_{m+1} p(l_{m+1})
\theta\bigg( 
S_{m}< { L}^{*}  \;\wedge S_{m}  > { L}^{*} - l_{m+1}
 \bigg)
\underbrace{
 \int \bigg(\prod_{\nu=1}^{m} d l_{\nu}\bigg)
\delta(S_{m}=\sum_{\nu=1}^{m} l_{\nu})
\prod_{\nu}^{m} p(l_{\nu})\
}_{\color{blue} = p(S_{m}\CS \BK)}\notag \\
&=\int_{0}^{\infty} dl_{m+1} p(l_{m+1})   \int_{{ L}^{*}-l_{m+1}}^{{ L}^{*}} 
dS_{m} p(S_{m}\CS \BK)\;.
\end{align}
The $l_{\nu}$ are iid and therefore we rename $l_{m+1}$ generically to $l_{1}$. Hence,
\begin{align}
P(m \CS \BK) &=
\int_{0}^{\infty} dl_{1} p(l_{1})   \int_{{ L}^{*}-l_{1}}^{{ L}^{*}} 
dS_{m} \;p(S_{m}\CS \BK)\;.
\end{align}
Now $S_{m}$ is a sum of a very large number of iid random variables and its PDF therefore follows 
from the central limit theorem
\begin{align}
p(S_{m}) &= \frac{1}{\sqrt{2\pi m\sigma^{2}}}e^{-\frac{(S_{m}-m \langle l_{1} \rangle)^2}{2m\sigma^{2}}} \notag, \\
\sigma^{2} &=  \langle \big(\Delta l_{1}\big)^{2} \rangle\;.
\end{align}
This can be cast into a Gaussian with respect to $m$
\begin{align}
p(S_{m}) 
&=\frac{1}{\langle l_{1} \rangle}{\cal N}(m\CS m_{0}(S_{m}), m_{0}(S_{m})\tilde\sigma^{2})\;,
\intertext{with}
m_{0}(S_{m})&=\frac{S_{m}}{\langle l_{1} \rangle}\;,\qquad
\tilde\sigma^{2} =  \frac{\langle \big(\Delta l_{1}\big)^{2} \rangle}{\langle l_{1} \rangle^{2}}\;.
\end{align}
For the transformation  we used that the peak of this Gaussian is at $m_{0}\gg 1$, which allows to replace 
$m \tilde\sigma^{2}$  by $m_{0} \tilde\sigma^{2}$.
Then we can easily determine the moments $\langle m^{\gamma} \rangle$
\begin{align}
\langle m^{\gamma} \rangle
&= \frac{1}{\langle l_{1} \rangle}\int_{0}^{\infty} dl_{1} p(l_{1})\;\int_{{ L}^{*}-l_{1}}^{{ L}^{*}} dS_{m}
\int_{m=0}^{\infty} dm \;
{\cal  N} \big( m\CS m_{0}(S_{m}),m_{0}(S_{m})\tilde \sigma \big)
\;m^{\gamma}\;.
\end{align}
We have replaced the sum over $m$ by an integral,
since  $p(S_{m})$ as function of $m$ is a slowly varying. 
The mean is
\begin{align}
\langle m \rangle
&= \frac{1}{\langle l_{1} \rangle}\int_{0}^{\infty} dl_{1} p(l_{1})\;\int_{{ L}^{*}-l_{1}}^{{ L}^{*}} dS_{m}\;
\;m_{0}(S_{m})\nonumber\\
&= \frac{{ L}^{*}}{\langle l_{1} \rangle} - \frac{1}{2}
\frac{\langle l_{1}^{2}\rangle }{\langle l_{1}\rangle^{2} }\;.\label{app:avg:m}
\end{align}
Similarly we obtain for the second moment
\begin{align}
\langle m^{2} \rangle
&=  \tilde\sigma^{2} \langle m \rangle
+
\frac{\big({ L}^{*}\big)^{2}}{\langle l_{1} \rangle^{2}}  -
{ L}^{*} \frac{\big\langle l_{1}^{2}\big\rangle}{\langle l_{1} \rangle^{3}}
+ \frac{1}{3}\frac{\langle l_{1}^{3}\rangle}{\langle l_{1} \rangle^{3}}\label{app:avg:mq}
\end{align}
Subtracting $\langle m \rangle^{2}$ based on \eq{app:avg:m} finally yields the variance  
\begin{align}
\langle \big(\Delta m\big)^{2} \rangle 
&=  \tilde\sigma^{2} \langle m \rangle
+ \frac{1}{3}\frac{\langle l_{1}^{3}\rangle}{\langle l_{1} \rangle^{3}}
-\frac{1}{4}
\bigg(\frac{\langle l_{1}^{2}\rangle }{\langle l_{1}\rangle^{2} }\bigg)^{2}\;.
\label{app:var:m}
\end{align}
In \eq{app:avg:m} and \eq{app:var:m} the first term is of order $N$ while the  others are of order 1 and hence negligible.
In summary, by ignoring the $O(1)$ terms, we have
\begin{align}\label{eq:n:max}
\langle n_\text{max} \rangle  &= \frac{{ L}^{*}}{\langle l_{1} \rangle},\\
\langle \big(\Delta n_\text{max}\big)^{2} \rangle 
&= \langle n_\text{max} \rangle\;  
 \frac{\langle \big(\Delta l_{1}\big)^{2} \rangle}{\langle l_{1} \rangle^{2}}.
\end{align}
and using $L^*\propto N$ and \eq{eq:fixedseq10} produces
\begin{align}
    \langle n_\text{max} \rangle \propto \frac{NK}{k} \;.
\end{align}

\subsection{Distribution of $X_\text{max}$}\label{app:X:max}
We are interested in the properties of
\begin{align}
-\ln\big( X_{n_\text{max}} \big) &= \sum_{\nu=1}^{n_\text{max}} l_{\nu}\;.
\end{align}
We recall that  the individual $l_{\nu}$ are iid and $n_\text{max}$ is an additional independent random variable. Consequently, for the mean we have two averages, one over the set of shrinkage factors and one over $n_\text{max}$. The results for mean and variance are
\begin{align}
\bigg\langle -\ln\big( X_{n_\text{max}} \big)\bigg\rangle 
&= \langle n_\text{max}\rangle\; \langle l_{1}\rangle\;,\label{app:avg:X:max}\\
\bigg\langle \bigg(\Delta \ln\big( X_{n_\text{max}} \big)\bigg)^{2}\bigg\rangle 
&=2\;\langle n_\text{max}\rangle\;\langle \big(\Delta l_{1}\big)^{2}\rangle\;.
\label{app:var:X:max}
\end{align}
The proof for the variance is as follows
\begin{align}
\bigg\langle \bigg(\Delta\ln\big( X_{n_\text{max}} \big)\bigg)^{2}\bigg\rangle &= \sum_{n_\text{max}=0}^{\infty} P(n_\text{max})\sum_{\nu,\nu'=1}^{n_\text{max}} \langle l_{\nu}l_{\nu'}\rangle \notag \\
&= \sum_{n_\text{max}=0}^{\infty} P(n_\text{max})
\bigg(
\sum_{\nu}^{n_\text{max}} \langle l_{\nu}	^{2}\rangle
+
\sum_{\nu\ne \nu'}^{n_\text{max}} \langle l_{\nu}\rangle\langle l_{\nu'}\rangle
\bigg)\notag \\
&= \sum_{n_\text{max}=0}^{\infty} P(n_\text{max})
\bigg(
n_\text{max}\;\langle l_{1}^{2}\rangle
+
n_\text{max}\big( n_\text{max}-1 \big)\;\langle l_{1}\rangle^{2}
\bigg)\notag \\
&= 
\langle n_\text{max}\rangle\;\langle \big(\Delta l_{1}\big)^{2}\rangle
+
\langle n^{2}_\text{max}\rangle\;\langle l_{1}\rangle^{2}\;.
\end{align} 
Subtracting the mean squared yields the  expression for the variance given in \eq{app:var:X:max}.

\subsection{High temperature limit  of $\avg{|{\cal M}|}_{\beta}$\label{app:hightempM}}
Here we determine the high temperature limit of  $\avg{|{\cal M}|}$. We start out from the definition 
\begin{align}
\avg{|{\cal M}|}_{\beta=0} &= \sum_{\{\vv s \}}  \boldsymbol{\bigg\vert}\sum_{i} e^{i\frac{2\pi}{q} s_{i}}\boldsymbol{\bigg\vert}\; P(\{s\}\CS N, \beta=0)\notag\\
&= \sum_{\{\vv n \}}  \boldsymbol{\bigg\vert}\sum_{s=1}^{q} e^{i\frac{2\pi}{q} s }n_{s}\boldsymbol{\bigg\vert}\; P(\{n\}\CS N,\beta=0)\;,\label{app:M:a}
\end{align}
here $n_{s}$ is the number site at which the spin has the value $s$. As the spins are independently and identically distributed, the probability $P(\{n\}\CS N,\beta=0)$ is multinomial 
\begin{align*}
P(\{n\} \CS N,\beta=0) &= \delta\big( \sum_{s}n_{s}-N \big)\;\frac{N}{\prod_{s=1}^{q}n_{s}!} q^{-N}\;,
\end{align*}
with mean $\avg{n_{s}}=\frac{N}{q}$ and variance $\avg{(\Delta n_{s})^{2}} = \frac{N (q-1)}{q^{2}}$.
In  \eq{app:M:a} we introduce $n_{s} = \avg{n_{s}} + \Delta n_{s}$, resulting in  
\begin{align}
\avg{|{\cal M}|}_{\beta=0} 
&=  \sum_{\{\vv n \}}  \boldsymbol{\bigg\vert}
\frac{N}{q} \underbrace{
\sum_{s=1}^{q} e^{i\frac{2\pi}{q} s }
}_{\color{blue} = 0}
+
\sum_{s=1}^{q} e^{i\frac{2\pi}{q} s } \Delta n_{s}\boldsymbol{\bigg\vert}
P(\{\Delta n\}\CS N,\beta=0)
\\
&\le  \sum_{\{\vv n \}} 
\sum_{s=1}^{q}  \boldsymbol{\big\vert} \Delta n_{s}\boldsymbol{\big\vert}
P(\{\Delta n\}\CS N,\beta=0)\\
\\
&\le  q\;\sum_{\{\vv n \}} 
 \boldsymbol{\big\vert} n_{1} - \frac{N}{q}\boldsymbol{\big\vert}
P(n_{1}\CS N,\beta=0)\;.
\end{align}
For the case $q=2$ the equal sign applies. The marginal probability distribution of
the multinomial is the binomial for which we can invoke  the Moivre-Laplace theorem, as we are interested in $N\gg 1$. Then
we obtain
\begin{align}\label{eq:high:T:mod:magn}
\avg{|{\cal M}|}_{\beta=0}
&\le \frac{q}{\sqrt{2\pi\sigma^{2}}}\;\int_{0}^{N} 
\boldsymbol{\big\vert} n - \avg{n}\boldsymbol{\big\vert}  \;
e^{-\frac{(n-\avg{n})^{2}}{2 \sigma^{2}}}\;dn \notag \\
&\le \frac{2q}{\sqrt{2\pi\sigma^{2}}}\;\int_{0}^{\infty} x  \;e^{-\frac{x^{2}}{2 \sigma^{2}}}\;dx 
= \frac{2q\sigma}{\sqrt{2\pi}}\;.
\end{align}
Inserting the value of the variance  $\sigma^{2} = \frac{N}{q^{2}}(q-1)$
yields
\begin{align}\label{app:M:limit}
\frac{\avg{|{{\cal M}}|}_{\beta=0}}{N}
&\le  \sqrt{\frac{2 (q-1)}{\pi N}}\underset{N\to \infty} {\longrightarrow}0
\end{align}

\subsection{Expectation values\label{app:expectationValue:Potts}}
Now we repeat the steps of appendix \ref{app:NESA:Z} for the evaluation of an expectation value
\begin{align}
\langle O \rangle &= \frac{1}{Z}\;\SumInt{x} L(\vv x) \pi(\vv x)\;O(\vv x)\;.
\end{align}
We find
\begin{align}
\langle O \rangle &= \frac{1}{Z}\int d\lambda\; \lambda\;\SumInt{\vv x}  \; O(\vv x) \pi(\vv x) \delta(L(\vv x)-\lambda) \notag \\
 &= \frac{1}{Z}\int d\lambda\; \lambda\;
 \underbrace{
\frac{\SumInt{\vv x} \; O(\vv x) \pi(\vv x) \delta(L(\vv x)-\lambda)}{\SumInt{\vv x} \; \pi(\vv x) \delta(L(\vv x)-\lambda)}
}_{\color{blue} := \langle O \rangle_{\lambda}}\;
\underbrace{
\SumInt{\vv x}  \; \pi(\vv x) \delta(L(\vv x)-\lambda)
}_{\color{blue} = -\frac{d X(\lambda)}{d\lambda}} \notag \\
 &= \frac{1}{Z}\int \; {\cal L}(X)\; \langle O \rangle_{{\cal L}(X)} \;dX\;.
\end{align}
  The final result is therefore
\begin{align}\label{eq:avg}
\langle O \rangle&=\frac{ \int \; {\cal L}(X)\; \langle O \rangle_{{\cal L}(X)} \;dX}{ \int \; {\cal L}(X) \;dX} \;.
\end{align}

In case of the Potts model we have to replace the original spin degrees of freedom $\vv s$ by bond variables 
$\vv b$. The natural observables, however, are functions of the spins $O(\vv s)$.
\Eq{eq:avg} for the Potts model reads
\begin{align}\label{eq:}
\langle O \rangle&=\frac{1}{Z}\;
\sum_{\vv b}\int dm\; {\cal L}(\vv b,m) \langle O \rangle_{{\cal L}(\vv b,m)} \pi(\vv b,m)\;,
\end{align}
where $m$ stands for the auxiliary variable, introduced to lift the degeneracy.
Next we compute $\langle O \rangle_{{\cal L}(\vv b,m)}$. For observable that depend on the spins,  we have to reintroduce the spin degrees of freedom
\begin{align}\label{eq:}
\langle O \rangle_{{\cal L}(\vv b,m)} &= \frac{
\sum_{\vv b'}\int dm' 
\bigg(\sum_{\vv s} O(\vv s) P(\vv s\CS \vv b')\bigg)\;\pi(\vv b')\;\pi(m')\;\delta\big( {\cal L}(\vv b,m) - {\cal L}(\vv b',m') \big)
}{\sum_{\vv b'}\int dm' \;
\pi(\vv b')\;\pi(m')\;\delta\big( {\cal L}(\vv b,m) - {\cal L}(\vv b',m') \big)}
\end{align}
Next we use
\begin{align}\label{eq:}
\delta\bigg( {\cal L}(\vv b,m) - {\cal L}(\vv b',m') \bigg)
= \delta_{{\cal L}(\vv b),{\cal L}(\vv b')}\;\delta\big( m - m'\big)
\end{align}
and obtain
\begin{align}\label{eq:}
\langle O \rangle_{{\cal L}(\vv b,m)} &= \sum_{\vv b'}\;
\bigg(\sum_{\vv s} O(\vv s) P(\vv s\CS \vv b')\bigg)\;\frac{ \pi(\vv b')\;\delta_{{\cal L}(\vv b),{\cal L}(\vv b')}
}{\sum_{\vv b'}
\pi(\vv b')\delta_{L(\vv b),L(\vv b')}}\nonumber\\
&= \sum_{\vv b'}\;
\bigg(\sum_{\vv s} O(\vv s) P(\vv s\CS \vv b')\bigg)\;
\pi(\vv b'\CS {\cal L}(\vv b))\;,
\end{align}
where the latter quantity is the normalized prior, constrained by ${\cal L}(\vv b')={\cal L}(\vv b)$.
Since the observable is independent of the auxiliary variable $m$,
the expectation value $\langle O \rangle_{{\cal L}(\vv b,m)}$ is also independent of $m$.
We can use the algorithm, outlined in \Se\ref{ssec:nsbasic}, to  generate
samples according to the constrained prior $\pi(\vv b'\CS {\cal L}(\vv b))$. The sample size of such bond configurations shall be denoted by $N_{b}$.
As argued by Skilling \cite{skilling_nested_2006} and further developed in 
\cite{chopin_properties_2010}, instead of averaging over $\vv b'$, we can also use just the configuration $\vv b'$, corresponding to the walker with the minimum likelihood value $\lambda^{*}_{n}$, as it represents a valid sample. In the worst case, it merely leads to an
increased statistical error of the final result. It should , however, be pointed out that
the statistical error of this bond sampling is not taken into account by the uncertainty estimated
by sampling the prior masses. It can only be inferred from the sample with $N_{b}> 1$ or by repeated
nested sampling runs where $N_{\text{avg}}$ counts the repetitions.

For the inner average over the spins, we
need the conditional probability $P(\vv s\CS \vv b)$. In the context of the Swendsen Wang algorithm it  demands that all spins within one cluster have the same value, which otherwise is uniform, i.e.
\begin{align}\label{eq:}
P(\vv s\CS \vv b) &= \prod_{l=1}^{{\cal C}(\vv b)}
P({\cal S}_{l})\; \sum_{i\in {\cal C}_{l}}\;\delta_{s_{i},{\cal S}_{l}}\;,\\
P({\cal S}_{l}) &= \frac{1}{q}\;\theta\big( 1\le {\cal S}_{l}\le q \big).
\end{align}
Here ${\cal C}_{l}$ represents the set of sites belonging to cluster $l$. Also the inner sum over spins 
and its variance can be inferred from a sample of size $N_{S}$, say.

\section*{ACKNOWLEDGMENTS}

We acknowledge fruitful discussions with H. G. Evertz and U. von Toussaint
as well as the support of I. Murray.

\section*{References}

\bibliography{References}

\end{document}